\documentclass[a4paper,11pt]{article}
\usepackage{jcappub} % for details on the use of the package, please see the JINST-author-manual
\arxivnumber{2406.15832} % Only if you have one
\title{\boldmath Comparing sampling techniques to chart parameter space of 21 cm Global signal with Artificial Neural Networks}

\author[a,1]{Anshuman Tripathi,\note{Corresponding author.}}
\author[a,b]{Gursharanjit Kaur,}
 \author[a]{Abhirup Datta,}
\author[a,c]{Suman Majumdar}

\affiliation[a]{Department of Astronomy, Astrophysics and Space Engineering, Indian Institute of Tech-
nology Indore, Indore, India – 453552.}
\affiliation[b]{Center for Theoretical Physics, Polish Academy of Sciences, al. Lotników 32/46, 02-668 Warsaw, Poland}
\affiliation[c]{ Department of Physics, Blackett Laboratory, Imperial College, London SW7 2AZ, U. K}

\emailAdd{anshumantripathi85@gmail.com}
\abstract{Understanding the first billion years of the universe requires studying two critical epochs: the Epoch of Reionization (EoR) and Cosmic Dawn (CD). However, due to limited data, the properties of the Intergalactic Medium (IGM) during these periods remain poorly understood, leading to a vast parameter space for the global 21cm signal. Training an Artificial Neural Network (ANN) with a narrowly defined parameter space can result in biased inferences. To mitigate this, the training dataset must be uniformly drawn from the entire parameter space to cover all possible signal realizations. However, drawing all possible realizations is computationally challenging, necessitating the sampling of a representative subset of this space. This study aims to identify optimal sampling techniques for the extensive dimensionality and volume of the 21cm signal parameter space. The optimally sampled training set will be used to train the ANN to infer from the global signal experiment. We investigate three sampling techniques: random, Latin hypercube (stratified), and Hammersley sequence (quasi-Monte Carlo) sampling, and compare their outcomes. Our findings reveal that sufficient samples must be drawn for robust and accurate ANN model training, regardless of the sampling technique employed. The required sample size depends primarily on two factors: the complexity of the data and the number of free parameters. More free parameters necessitate drawing more realizations. Among the sampling techniques utilized, we find that ANN models trained with Hammersley sequence sampling demonstrate greater robustness compared to those trained with Latin hypercube and Random sampling.}


\begin{document}
\maketitle
\flushbottom

\section{Introduction}
\label{sec:intro}
The $\Lambda$CDM (Lambda Cold Dark Matter) cosmological model asserts that the initial structures in the universe formed during the Cosmic Dawn as hydrogen gas collapsed under the effect of gravity. The ultraviolet (UV) photons emitted by the first luminous sources initiated the ionization of the surrounding intergalactic medium (IGM), causing the last transition phase in the universe's evolution. This phase transition period is called the Epoch of Reionization (EoR) \citep{Furlanetto2006, fan2006, morales2010, pritchard201221}. Due to the lack of observational data at these high redshifts, the characteristics and timeline of the EoR remain poorly constrained. The HI 21cm signal stands out as a promising probe for exploring these uncharted redshift ranges. This signal originates from the hyperfine splitting of the ground state of the hydrogen atom, which arises from the spin alignments of the proton and electron \citep{Field1958, Field1959a, Field1959b}. The "spin-flip" transition responsible for the signal is inherently forbidden. Nevertheless, an abundance of hydrogen in the universe makes it an important astrophysical probe \citep{2013ASSL..396...45Z}. 

However, there are several challenges in observing the signal. The 21 cm signal is overshadowed by the foregrounds, $10^{4}$ times brighter than the signal in the redshifted frequency range. The foregrounds are predominantly due to galactic synchrotron emission. These foregrounds, along with the ionospheric effects and the instrument response to the observation, pose a significant challenge to the detectability. Two distinct experimental techniques are used for observing these faint cosmological signals. One approach involves single-radiometer, as seen in experiments like  Experiment to Detect the Global Epoch of Reionization Signature (EDGES, \citep{Bowman2018}), Shaped Antenna Measurement of the Background Radio Spectrum (SARAS, \citep{saras, Saurabh2022}), Radio Experiment for the Analysis of Cosmic Hydrogen (REACH, \citep{reach}) and  Large-Aperture Experiment to Detect the Dark Ages (LEDA, \citep{leda}). Alternatively, interferometers such as the Giant Meterwave Radio Telescope (GMRT, \citep{gmrt1991}), Hydrogen Epoch of Reionization Array (HERA, \citep{HERA2017}), Low-Frequency Array (LOFAR \citep{lofar2013}), and the upcoming Square Kilometer Array (SKA, \citep{ska2015}) are utilized. Recently, EDGES has reported a possible detection of the sky-averaged global 21 cm signal with an unexpectedly deep absorption trough at 78 MHz \citep{Saurabh2022}. The depth is more than twice what the standard cosmology model predicted. If true, the signal can give new insights into the physics of the reionization era. However, another independent experiment, SARAS, has challenged this detection, suggesting that the anomalous result might be due to uncalibrated systematics \citep{Bowman2018}. Thus, it emphasizes that correctly modelling and removing the corruption from the signal is essential.

Over recent years, machine learning (ML) algorithms have gained extensive popularity in signal modelling and parameter estimation. For signal modelling, approaches such as those by \citep{cohen2020emulating, schmit2018emulation, globalemu, VAE, tiwari2022, jennings2019evaluating} leverage Artificial Neural Networks (ANN) to model the 21cm signal across various aspects. Simultaneously, \citep{2018Shimabukuro, choudhury2020extracting, choudhury2021using, 2022Choudhury, Tripathi2024} employ ANN algorithms to extract parameters linked to the signal. Additionally, simulation-based inference (SBI) techniques, leveraging deep learning, are increasingly preferred for astrophysical inference from cosmic 21 cm signals, particularly for retrieving posteriors of astrophysical parameters via likelihood-free Bayesian inference directly from power spectrum and light-cone image analysis \citep{2022_a_Zhao_3D_lightcone, 2022_b_Zhao_PS, 2023Prelogovi, 2023Saxena, 2024Greig}. \citep{2023Zhao} introduced the Solid Harmonic Wavelet Scattering Transform (WST) and "3D ScatterNet" to enhance the inference of astrophysical parameters. Apart from the ML algorithm, several traditional methods are used to remove the signal and associated parameters \citep{harker2012mcmc, Rapetti_2020, Tauscher2021, Bevins_2021}. Currently, due to limited observational constraints on intergalactic medium (IGM) properties, there is no singular and firmly established set of quantities to parameterize the underlying astrophysical processes shaping the signal. Several potential parameters have been suggested in various proposed models of the signal\citep{2015Greig, 2017Greig, cohen2017charting, 2017Semelin}.

Previously, \cite{choudhury2020extracting, choudhury2021using, Tripathi2024} successfully developed and presented an ANN model capable of extracting astrophysical parameters of the 21 cm signal from mock observation data sets. These models considered limited sets of possible signal combinations, incorporating the effects of foregrounds, ionosphere, instruments, and noise. To develop a robust ANN model for foreground removal, it is crucial to consider all possible varieties of signals in the training set \citep{cohen2017charting}. However, considering all the possible signal combinations will be computationally expensive. To address this issue, in this study, we create a sub-sample from the entire signal parameter space, ensuring it represents the overall parameter space using various sampling methods. Nonetheless, there are no straightforward rules to determine the optimal sampling technique and identify the minimum number of samples required for training the ANN model in a robust and accurate manner, which poses uncertainties. Addressing these questions requires a more in-depth exploration. 

To address these questions, this study explores three distinct sampling methods—Random (Rand) sampling, Latin hypercube sampling (LHS), and Hammersley sequence sampling (HSS)—to comprehensively map the parameter space and generate various global 21cm signal types. We also analyzed the minimum number of samples required to train the ANN effectively, ensuring its robustness across different signal types. Furthermore, our investigation aims to understand the efficiency of these sampling algorithms and determine the minimum training data size needed to achieve consistent accuracy as the parameter space dimensionality and dataset complexity increase, particularly with the addition of foreground and thermal noise to the global 21cm signal. Additionally, we conducted generalizability tests by generating multiple training datasets through repeated parameter space sampling. The ANN was trained multiple times with these datasets and subsequently tested with unknown datasets generated using various sampling methods. This approach allows us to examine the consistency of the sampling algorithms in effectively covering the parameter space and to understand the clustering issues associated with these sampling methods by examining the consistency of the ANN's predictions.

This paper is structured as follows: Section \ref{21cm_sig} outlines the observable aspects of the HI 21cm Signal. Section \ref{obs_challenge} delves into the observational challenges, while Section \ref{siggen} describes the methods for simulating the global 21cm signal. Section \ref{sampling_tech} covers sampling techniques, and Section \ref{ANN_overview} provides a basic overview of ANN. Section \ref{Data_pre} discusses the training and test datasets for the ANN, and Section \ref{result} outlines the results, including a discussion of the ANN predictions. The final section, Section \ref{summary_discussion}, comprises the summary and overall discussions.

\section{HI 21cm Signal}\label{21cm_sig}
The 21cm signal arises because of hyperfine splitting of 1S ground level of hydrogen atom due to the interaction of magnetic moments of electrons and protons. Commonly, this transition is known as the spin-flip transition, where the spin transition from parallel to anti-parallel takes place. This transition results in the spontaneous emission of 21cm photon.

A single dish experiment measures the brightness temperature of the signal $T_{b}$ in contrast to the background temperature of Cosmic Microwave Background (CMB), $T_{CMB}$; this is called the differential brightness temperature:
\begin{align}
\delta T_{b} \approx 27(1 - x_{HI})\left (\frac{\Omega _{b}h^2}{0.023} \right) \left(\frac{0.15}{\Omega _{m,0}}\frac{1+z}{10}\right) ^{1/2} \left(1 - \frac{T_{CMB}(z)}{T_{s}}\right) \end{align}
where $x_{HI}$ is the neutral fraction of hydrogen, $\Omega_{b}$ and $\Omega_{m}$ are the
baryon and total matter density, respectively, in units of the critical density, $H(z)$ is the Hubble parameter at redshift $z$ and, $T_{s}$ is the spin temperature of neutral hydrogen.

The spin temperature, which is the relative populations of hydrogen atoms in the two spin states, is decided by competition between 3 processes and their corresponding physical quantities: (1) absorption of CMB photons and stimulated emission; $T_{CMB}$, the temperature of CMB, (2) collisions with other hydrogen atoms (H-H), free electrons (H-e), free hydrogen nuclei (H-p); $T_{k}$, Kinetic gas temperature of IGM, and (3) scattering of Lyman- $\alpha$ photons, $T_{\alpha}$, colour temperature for Wouthuysen–Field effect. The spin temperature, $T_{s}$ is calculated as  \citep{Field1959a}\citep{10.1088/2514-3433/ab4a73ch1}:  
\begin{align}
T_{s}^{-1} = \frac{T_{CMB}^{-1} + x_{k} T_{k}^{-1} + x_{\alpha} T_{\alpha}^{-1} }{1 + x_{k} + x_{\alpha}}
\label{eq:couple}
\end{align} 
Here, $x_{k}, x_{\alpha}$ are collisional and Lyman- $\alpha$ coupling coefficients. Thus, the global signal evolves over the redshift range as a function of the properties of IGM.

\section{Observational Challenges}\label{obs_challenge}
Observing the redshifted 21cm signal poses challenges due to various observational obstacles, including bright foregrounds, ionospheric effects, beam chromaticity, thermal noise, and radio frequency interference (RFI). In this study, our primary focus is on addressing two specific challenges: the impact of foreground and thermal noise, while simulating observations to construct the training datasets.
\subsection{Foregrounds}
About 70\% of the foregrounds obscuring the 21 cm signal come from galactic synchrotron sources, and the rest are free-free emissions and thermal dust emissions. Extragalactic foregrounds are primarily caused by radio emission from star-forming galaxies. There are two ways to deal with the foreground: avoiding and removing it. The former can be applied for interferometric observations but not for the global 21cm signal. For global 21cm experiments, the foregrounds must be modelled and removed. The foregrounds are spectrally smooth compared to the global 21cm signal. The high coherence of the diffuse galactic foregrounds across frequency compared to the signal can be used for foreground subtraction \citep{liu2009improved}. Due to their spectral smoothness, the foregrounds can be modelled as a low-order polynomial \citep{pritchard201221}. In this study, we simulated the diffuse foreground using a third-order polynomial in $log(\nu)-log(T)$, as previously described by \citep{choudhury2020extracting, harker2015selection}. 
\begin{align}
\log(T_{FG}) =  \sum_{i = 0}^{n}a_{i}\left(\log \left(\dfrac{\nu}{\nu_{0}}\right)\right)
\end{align}
where $\nu_{0}$ = 80 MHz. The four foreground parameters, constants of the $log(\nu) - log(T)$ polynomial,  are varied around their inferred value by {\cite{harker2015selection}}, $a_{0} = \log(T_{0}) = 3.30955$; $a_{1} = -2.42096$; $a_{2} = -0.08062$; $a_{3} = 0.02898$.

\subsection{Thermal Noise}
The thermal noise, denoted as $n(\nu)$, in the observed spectrum can be expressed using the ideal radiometer equation in the following manner:
\begin{equation}
\begin{split}
{n(\nu) \approx \frac{T_{sys}(\nu)}{\sqrt{\delta \nu  \cdot \tau}}},
\end{split}
\end{equation}
In this context, $\rm T_{sys}(\nu)$ represents the system temperature, $\delta \nu$ is the observational bandwidth, and $\tau$ denotes the observation time.

\section{Models for the Global 21cm Signal}\label{siggen}
To simulate the global 21cm signal, we employed two distinct models. The first model is based on a parametrized model, while the other utilizes a semi-numerical astrophysical approach. Further details are provided below:

\subsection{\emph{tanh} Parametrization Model}
We have used the tanh model to simulate the global 21 cm signal during the cosmic dawn (CD) and the EoR using ARES \citep{2014MNRAS.443.1211M}. The model uses simple parametric forms for the Lyman  $\alpha$ background, IGM temperature, and re-ionization histories \citep{2016MNRAS.455.3829H}. This method models the signal using IGM properties like the strength of Lyman $\alpha$ coupling, $J_{\alpha}(z)$; the temperature of the IGM, $T(z)$; and ionization fraction, ${\overline{X}_{i}}$, but does not take into account the source properties. Each of these quantities is evolved as:
\begin{equation}
 A(z) = \frac{A_{ref}}{2}\left(1 + tanh\left(\frac{z_{0} - z}{\Delta z}\right)\right)
 \label{tanh_equation}
\end{equation}
Here $A_{ref}$ is the step height, $z_{0}$ is the pivot redshift, and $\Delta z$ is the width. The quantities become zero at high redshift, turn on for a redshift interval $\Delta z$ around a $z_{0}$, the central redshift, and achieve maximum $A_{ref}$ saturation at low redshift. 

The step height, $A_{ref}$, in units of  $10^{-21}$ $\rm ergs^{-1} cm^{-2}  Hz^{-1} sr^{-1}$  corresponding to the Lyman $\alpha$ ($Ly \alpha$) background is $J_{ref}$. It saturates at low redshift, with a value of 11.69, as determined by \cite{2016MNRAS.455.3829H} using MCMC parameter estimation. In our study, we treat it as a constant, as done in our work. The redshift interval and pivot redshift for $Ly\alpha$ background tanh parametrization are $J_{dz}$ and $J_{z0}$ respectively. Similarly, for X-ray heating, the temperature of IGM, $T(z)$,   $T_{dz}$ and $T_{z0}$ denotes redshift interval and central redshift in units of Kelvin. The amplitude, $ A_{ref}$,  corresponding to IGM temperature, represented by $T_{ref}$, saturates at around 1000K. For the ionization fraction, ${\overline{X}_{i}}$, the natural value for step height is unity. The redshift interval $\Delta z$ and the pivot redshift $z_{0}$ over which ionization takes place are given by $X_{dz}$ and $X_{z0}$, respectively. The values for these parameters inferred by \cite{2016MNRAS.455.3829H} are: $J_{dz}$  = 3.31, $J_{z0}$ = 18.54,  $T_{dz}$  = 2.82,  $T_{z0}$ = 9.77, $X_{dz}$ = 2.83, $X_{z0}$ = 8.68.

\subsection{Astrophysical Model}
We used a semi-numerical model which simulates the global signal based on the evolution of the properties of the IGM during the EoR, described in \citep{10.1093/mnras/stz1444}. The astrophysical input parameters to simulate the global 21 cm signal from the Cosmic Dawn are the following:
\begin{enumerate}
\item Star formation efficiency, $f_{\ast}$: A high value of star-forming efficiency implies earlier cosmic heating and a shallower absorption feature;
\item The escape fraction of the ionizing photons into the IGM, $f_{esc}$: The number of ionizing photons able to reach the IGM decides the duration of the reionization;
\item  X-ray heating efficiency, $f_{X}$: The efficiency of X-ray sources to heat the IGM decides the depth of the absorption trough. For high values of  $f_{X}$, the shallower absorption feature is shifted to higher redshifts; 
\item Number of Ly $\alpha$ photons produced per baryon, $N_{\alpha}$: The Lyman $\alpha$ background emission depends upon the metallicity of the stars;

\begin{figure*}
    \centering
    {\includegraphics[width=0.45\textwidth]{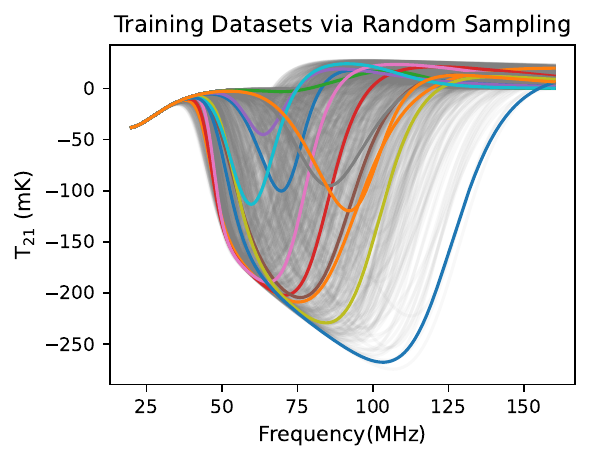}}
    {\includegraphics[width=0.45\textwidth]{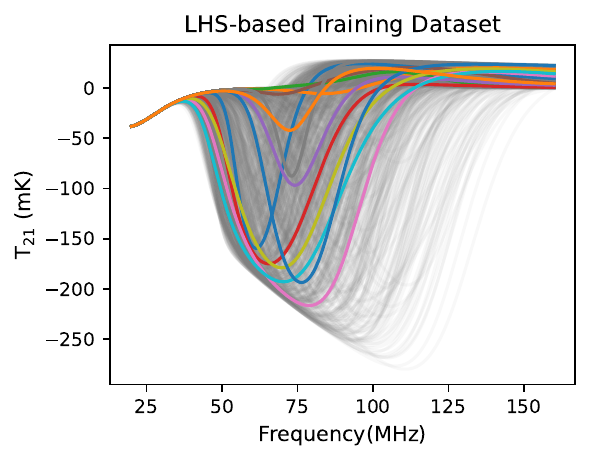}}
    {\includegraphics[width=0.45\textwidth]{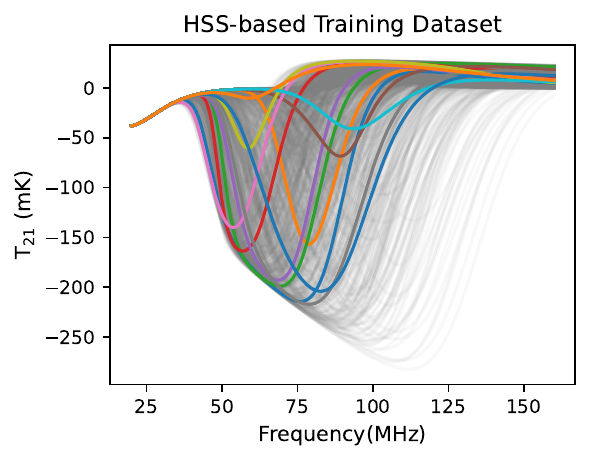}}
    {\includegraphics[width=0.45\textwidth]{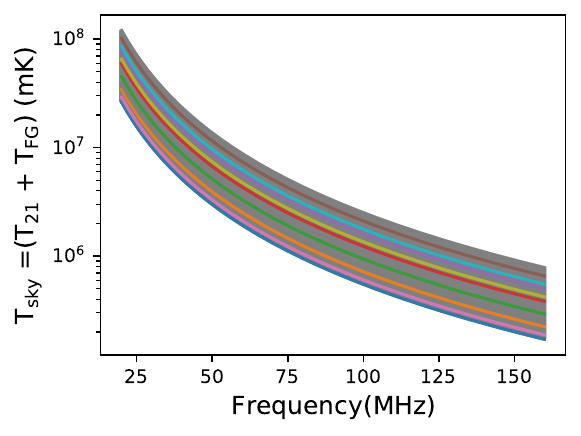}}
    \caption{Each subplot illustrates training datasets for the global 21cm signal, constructed using Random, Latin hypercube, and Hammersley sequence sampling methods, respectively, for the parametrized  signal. The final subplot depicts training datasets with added foreground and thermal noise. Signal subsets are highlighted in color, while the remaining sets are displayed in gray as the background.}
   
    \label{Fig1}
\end{figure*}

\begin{figure*}
    \centering
    {\includegraphics[width=0.45\textwidth]{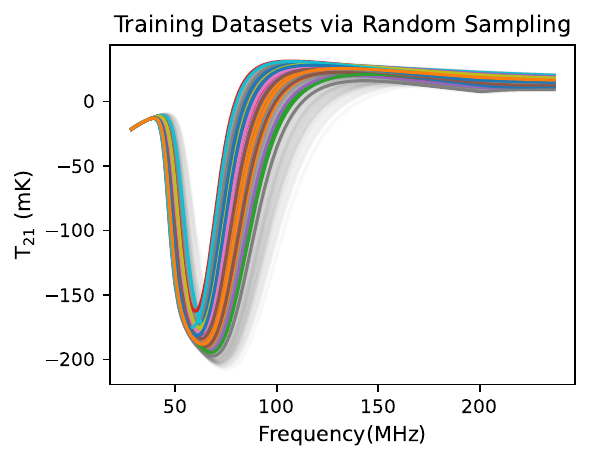}}
    {\includegraphics[width=0.45\textwidth]{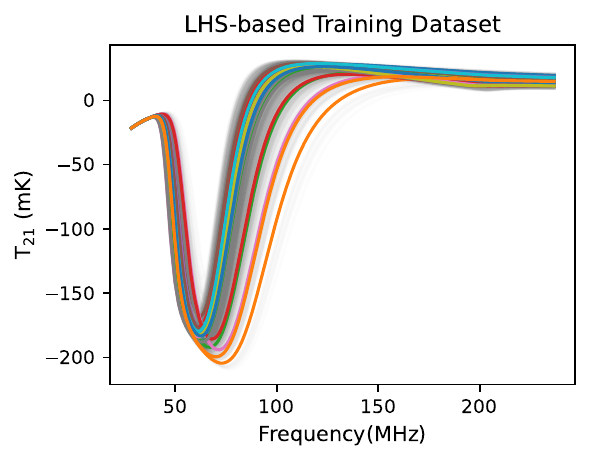}}
    {\includegraphics[width=0.45\textwidth]{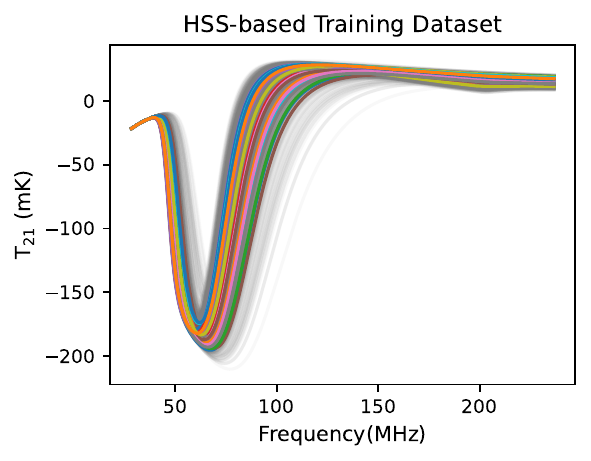}}
    {\includegraphics[width=0.45\textwidth]{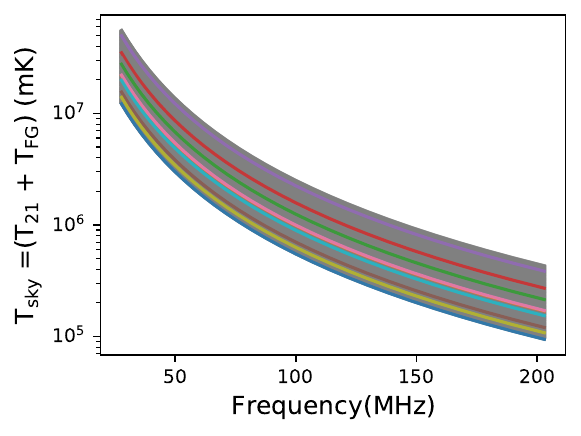}}
    \caption{Each subplot illustrates training datasets for the global 21cm signal, constructed using Random, Latin hypercube, and Hammersley sequence sampling methods, respectively, for the physical signal. The final subplot depicts training datasets with added foreground and thermal noise. Signal subsets are highlighted in color, while the remaining sets are displayed in gray as the background.}
   
    \label{Fig2}
\end{figure*}

\end{enumerate}
The parameters are described in detail in \citep{10.1093/mnras/stz1444, choudhury2021using}. Due to the limited observation for detecting the 21 cm signal from the Cosmic Dawn, the astrophysical parameters and the IGM properties still need to be determined. Some of the astrophysical parameters are poorly constrained, like the optical depth of the CMB by Planck data {\citep{2016A&A...596A.108P}}. Thus, the astrophysical parameters can have any value within an extensive range. For this work, we have assumed the following ranges for the parameters to simulate the different realization of the global 21cm as tabulated in Tab. \ref{tab:1}.

\section{Sampling the parameter space}\label{sampling_tech}
To train artificial neural networks (ANNs) or any machine learning model for parameter extraction from the 21 cm signal, a comprehensive training set is required, ideally covering all possible signal templates. As described earlier, the challenge is the computational power required to construct such a gallery, given that the parameters governing the signal theoretically span large ranges. The problem can be tackled by considering a sub-sample of the parameter space that optimally covers the range of all parameters. For instance, Cohen 2017 \cite{cohen2017charting} computed 193 signal realizations for different sets of astrophysical parameters. For this work, we consider and compare three different sampling techniques to optimally sample the parameter space of the global 21 cm signal:
\subsection{Random sampling}
Random sampling is the commonly used sampling to ensure an unbiased selection of points in the parameter space. Each parameter is assumed to have a uniform probability distribution. This implies that each possible value within the parameter space has an equal likelihood of being selected during the sampling process. 
\subsection{Latin hypercube sampling} 
Latin hypercube sampling (LHS) is a type of stratified sampling {\citep{10.2307/1268522}}. In LHS, the range of each input parameter is divided into N intervals having equal marginal probability $\dfrac{1}{N}$. The key idea is to ensure an even and representative sampling of the entire parameter space. For each parameter, N samples are drawn at random from each interval.  In the case of a Latin square with only two parameters $x$ and $y$, the $x_{i}$, $y_{i}$, for $i = 1,2,...,N$ are sampled independently. The samples taken from each parameter are matched at random as $(x_{i}, y_{j})$ for $i, j = 1,2,...,N$. A Latin hypercube extends this concept to higher-dimensional spaces. Simple Latin hypercube design and its variant have been commonly used in computer experiments designed for their space-filling properties. In cosmology, the use of LH algorithms has been explored for the construction of training sets for emulators \citep{2019JCAP...02..031R, jennings2019evaluating, schmit2018emulation, tiwari2022}.

LHS has certain merits over random sampling. Random sampling could lead to the clustering of points in specific regions of the parameter space, thereby creating an uneven representation across different regions. LHS ensures an even coverage of the whole parameter space by spreading the sample points over the entire range of each parameter. Furthermore, LHS can achieve this goal with a relatively small number of samples compared to random sampling. 
\subsection{Hammersley sequence sampling} 
Hammersley sequence sampling (HSS) is a sampling method developed by Kalagnanam and Diwekar \citep{10.2307/1271135} to address certain limitations associated with other sampling methods, particularly in the context of quasi-monte carlo methods and numerical integration. Low discrepancy sequences (LDS) like Hammersley sequences are one of the solutions to achieve a distribution of points with low discrepancy, where discrepancy is a metric for the deviation from a uniform distribution. The low discrepancy sequences have a deterministic structure as opposed to the stochastic nature of the random sampling, meaning that the sequence of points is fully determined by the number of points and the dimensionality of the space. Notably, in scenarios involving large-dimensional parameter spaces, the HSS demonstrates good uniformity over the Latin hypercube sampling \citep{kucherenko2015exploring}.

To construct a Hammersley sequence sample, n relatively prime numbers (integers that have no common divisors other than 1) are chosen: $p_{1}, p_{2}, .., p_{n}$. Each non-negative integer k can be expressed with a prime base p 
\begin{equation}
k = a_{0} + a_{1}p + a_{2}p^{2} + a_{3}p^{3} + ...
\end{equation}
where, $a_{i}$ is an integer in [0, p – 1]. Subsequently, the following expression is computed for each prime p:
\begin{equation}
\phi_{p}(k) = \frac{a_{0}}{p} + \frac{a_{1}}{p^{2}} + \frac{a_{2}}{p^{3}} + ...
\end{equation}
For d dimensions, the $k^{th}$ d-dimensional Hammersley’s point is $ (\frac{k}{n}
, \phi_{p_{1}}(k),  \phi_{p_{2}}(k), ..., \phi_{p_{d-1}}(k))$ for $k = 0, 1, 2, ..., n–1$ ,where $n$ is the total number of Hammersley’s points and $p_{1} < p_{2} < ... < p_{d–1}$. HSS finds its main applications in computer graphics and design optimization. This work is the first to employ this sampling method in the context of cosmology.

\section{Artificial Neural Networks}\label{ANN_overview}
Artificial neural networks (ANNs) are widely used supervised machine learning algorithms, with applications increasingly found in cosmology as well. Mathematically, for a given set of [$x_{i},y_{i}$], ANNs try to find a function, f(x), such that $y_{i} = f(x_{i})$ through a series of weighted summations. This operation occurs in a series of layers. The neurons are the building blocks of the ANN. The ANN contains n number of layers, with one input layer, one output layer, and n-2 hidden layers. The input values $x_{i}$ with D dimensions are given to the first layer with D nodes. The output from one layer $a_{i}$ goes to the next layer’s $j^{th}$ neuron as input $a_{j}$. The computation for the activation of the $j^th$ neuron in layer $l+1$ is given by:
\begin{align}
a^{l+1}_{j} = h\left(\sum_{i = 1}^{D}w_{ji}a^{l}_{i} + w_{j0}\right) 
\end{align}
Here, $h$ represents the activation function, $l$ denotes the layer index, and $a$ refers to the inputs for the corresponding layer. For l= 0, $a^{0}_{i} = x_{i}$ which represents the input values for the first layer. The terms $w_{ji}$ and $w_{j0}$ are weights and biases associated with the neuron set, respectively.  There is no activation function used in the output layer.

The aim is to train the model to minimize the loss function, i.e., finding the weight values that mimic the function. It involves updating the weights with an optimizer using back-propagation to reduce the loss at a rate called the learning rate. The default loss function in the training of ANNs, including in this work, is the Mean Squared Error (MSE), given by:
\begin{align}
\text{MSE} = \frac{1}{n} \sum_{i=1}^{n} (y_i - \hat{y}_i)^2 
\end{align}
Here, n is number of samples, $y_{i}$ is the true output label and $\hat{y}_i$ is the prediction after updating the weights, $w_{ji}$, and biases, $w_{j0}$, with every iterations of the ANNs. The prediction $\hat{y}_{j}$ for the $j^{th}$ neuron in the output layer is calculated as:
\begin{equation}
\hat{y}_j = \sum_{i=1}^{D} w_{ji} \cdot a^{L-1}_{i} + w_{j0}     
\end{equation}
Here, L is the index of the final layer. Most optimizers are based on the gradient descent method of loss optimization.

The number of layers, number of neurons in each layer, learning rate, and activation function are to predecided. These parameters are known as hyperparameters. We can search over a range of hyperparameters and use the ones giving minimum loss and maximum accuracy. In this work, Scikit-learn \footnote{\url{https://scikit-learn.org/stable/}} and Keras\footnote{\url{https://keras.io/}} are used for building an ANN. We optimized hyperparameters using trial and error to determine the best architecture for the ANNs. Details of the final architecture are provided in the following section. 

\subsection{Metric of accuracy}
For testing and validating the data set, RMSE and R-squared value is used as a measure to see the difference between predictions of input parameters made by the model for testing data and the actual value. 

If N samples are there for testing, and $\rm y_{true}, y_{pred}$ are the actual and predicted value of the parameter, $\rm {\overline{y}_{true}}$, mean of all actual values of the parameter. Then,
\begin{equation}
\text{RMSE}(y_{true}, y_{pred}) = \sqrt{\frac{1}{N}\sum (y_{true} - y_{pred})^2} 
\end{equation}
\begin{equation}
  R^{2}(y_{true}, y_{pred}) = 1 - \frac{\sum_{i=1}^{N}(y_{true} - y_{pred})^2}{\sum_{i=1}^{N}(\bar{y}_{true} - y_{true})^2}
\end{equation}

\section{Training and testing data set}\label{Data_pre}
In this work, we systematically explored parameter space using diverse sampling techniques to ensure comprehensive coverage. We generated a comprehensive array of global 21cm signals to enhance the robustness of our ANN model training. Additionally, we drew datasets of various sizes for each sampling method to determine the minimum sample size requirement for training the ANN more robustly and accurately. Our objective extended to evaluating the efficacy of various sampling methods in ANN training, described in section \ref{sampling_tech}. Furthermore, we also check the robustness of these trained ANN models, which are trained with the different sampled datasets by testing them with test datasets sampled with other sampling methods, to understand whether the prediction accuracy of the final trained ANN model is consistent with any random sets of the datasets or not. 

To perform the above operation, we used two different types of signal models, one parametrized and the other non-parametrized, based on the semi-numerical model, details described in section \ref{siggen}. In the parametrized model, we used 6 parameters to simulate the global 21cm signal, and for the non-parametrized model, we used 3 parameters to simulate the global 21cm signal. Incorporating these two distinct models not only introduced diversity in the signal parameters but also enabled the exploration of different dimensions within these parameters. Similarly, integrating foreground and noise into the dataset added complexity and increased the dimensionality of free parameters. We employed the same sampling methods used for the signal case to chart both signal and foreground parameters simultaneously.

\begin{table}
\centering

\begin{tabular}{|c|c|c|c|c|}
\hline
\multicolumn{1}{|c|}{} & \multicolumn{2}{|c|}{Parametrized} & \multicolumn{2}{|c|}{Physical}\\
\hline
& Parameters  & Ranges & Parameters  & Ranges \\ \hline
& $\rm J_{z0}$  & 9.27, 27.81    & $\rm f_{x}$ * $\rm f_{xh}$ & 0.0255, 7.9800  \\
& $\rm X_{z0}$ & 4.34, 13.02     & $\rm f_{star}$     &  0.0030, 0.0099\\   
& $\rm T_{z0}$ & 4.89, 14.65     & $\rm f_{esc}$      &  0.06, 0.19 \\
Signal & $\rm J_{dz}$ & 1.65, 4.96   & $\rm N_{\alpha}$  & 9000-800000      \\
& $\rm T_{dz}$ & 1.41, 4.23      &        &  \\
& $\rm X_{dz}$ & 1.42, 4.25     &    &     \\ \hline
& $\rm a_{0}$    &2.97, 3.64     & $\rm a_{0}$    &2.97, 3.64 \\
Foreground & $\rm a_{1}$  &-2.45,  -2.37    & $\rm a_{1}$  &-2.45,  -2.37\\
&$\rm a_{2}$   & -0.082, -0.079 & $\rm a_{2}$   & -0.082, -0.079 \\
&$\rm a_{3}$    & 0.027, 0.030 & $\rm a_{3}$    & 0.027, 0.030 \\ \hline
\end{tabular}

\caption{The range of parameters used to build the training dataset for the Parametrized and Non-parametrized  (Physical) cases of global 21cm signals and foregrounds.}
\label{tab:1}
\end{table}

\subsection{Signal only}
For this scenario, we simulated the training and testing datasets by sampling the parameter space of the global 21 cm signal, defined within a specific range of parameter values, using three distinct sampling methods. The signal simulated utilizing the physical model and tanh parametrization illustrated in Fig. \ref{Fig1} and \ref{Fig2}, respectively.

 We generated three datasets of varying sizes - 1000, 5000, and 10,000 samples - using given sampling methods. Subsequently, we utilized these datasets to train ANN models, aiming to analyze how the performance of the ANNs varies across these different dataset sizes. For the training of the ANN, we split these sampled datasets into a 7:3 ratio, allocating 70 \% for training and reserving the remaining 30 \% for network testing.

\subsection{Foreground and thermal noise corrupted signal}
In this scenario, our training dataset is created by exploring the parameter space of both the signal and foreground using the three given sampling methods. These sampled parameters are then utilized to construct the signal and foreground components. We utilized two different models to simulate the signal: the parametrized and physical models. We have simulated the foreground using a log-log polynomial model, as detailed in section 3. To simulate a realistic observational scenario spanning 1000 hours, we have added the thermal noise into the simulated signal and foreground datasets using the radiometer equation, resulting in the generation of our final training datasets. 
We generated three different dataset sizes: 10,000, 50,000, and 200,000 samples by drawing parameters from the specified ranges for the parametrized signal case. We created dataset sizes of 10,000, 50,000, and 100,000 samples for the physical case. Given the comparatively larger size of the datasets compared to scenarios where only the signal is considered, adhering to the common thumb rule, we have partitioned the datasets in a 9:1 ratio. Here, 90 \% is designated for training the ANN model, while 10\% is set aside for evaluating its performance, which should be sufficient. This split ensures an adequate representation of the variety of signals in the dataset, enhancing the accuracy and robustness of the model's predictions.

\section{Results}\label{result}
In this study, we trained the ANN model under two scenarios: one with the signal alone and another that included the signal, foreground, and thermal noise. These datasets were generated using three unique sampling methods: Random sampling, Latin hypercube sampling, and Hammersley sequence sampling. To ensure accurate and robust training, we employed varying dataset sizes, as elaborated below. We also employed two distinct types of global signals generated by two different models: one based on parametrized modelling and the other referred to as non-parametrized to showcase ANNs generalizability. The detailed result for the non-parametric signal is showcased in Appendix \ref{Non_parametric}.

\subsection{Signal only}

The model we use for training is constructed with Keras' Sequential API and comprises 1024 input neurons matching with 1024 frequency channels and three hidden layers with 64, 27 and 18 neurons, respectively; these hidden layers are activated by ReLU (Rectified Linear Unit)  and `tanh' activation function. The output layer has 6 neurons to predict the parametrized global 21cm signal parameters. The input training data sets are normalized and standardized with the 'MinMaxScaler' and `StandardScaler' functions, and corresponding parameters are normalized using `MinMaxScaler', available in Scikit-learn. We evaluated the performance of the trained ANN using a test dataset and assessed its prediction accuracy by calculating the $ \rm R^{2}$ and the RMSE. 

\begin{figure*}
    \centering
    {\includegraphics[width=1.0\textwidth]{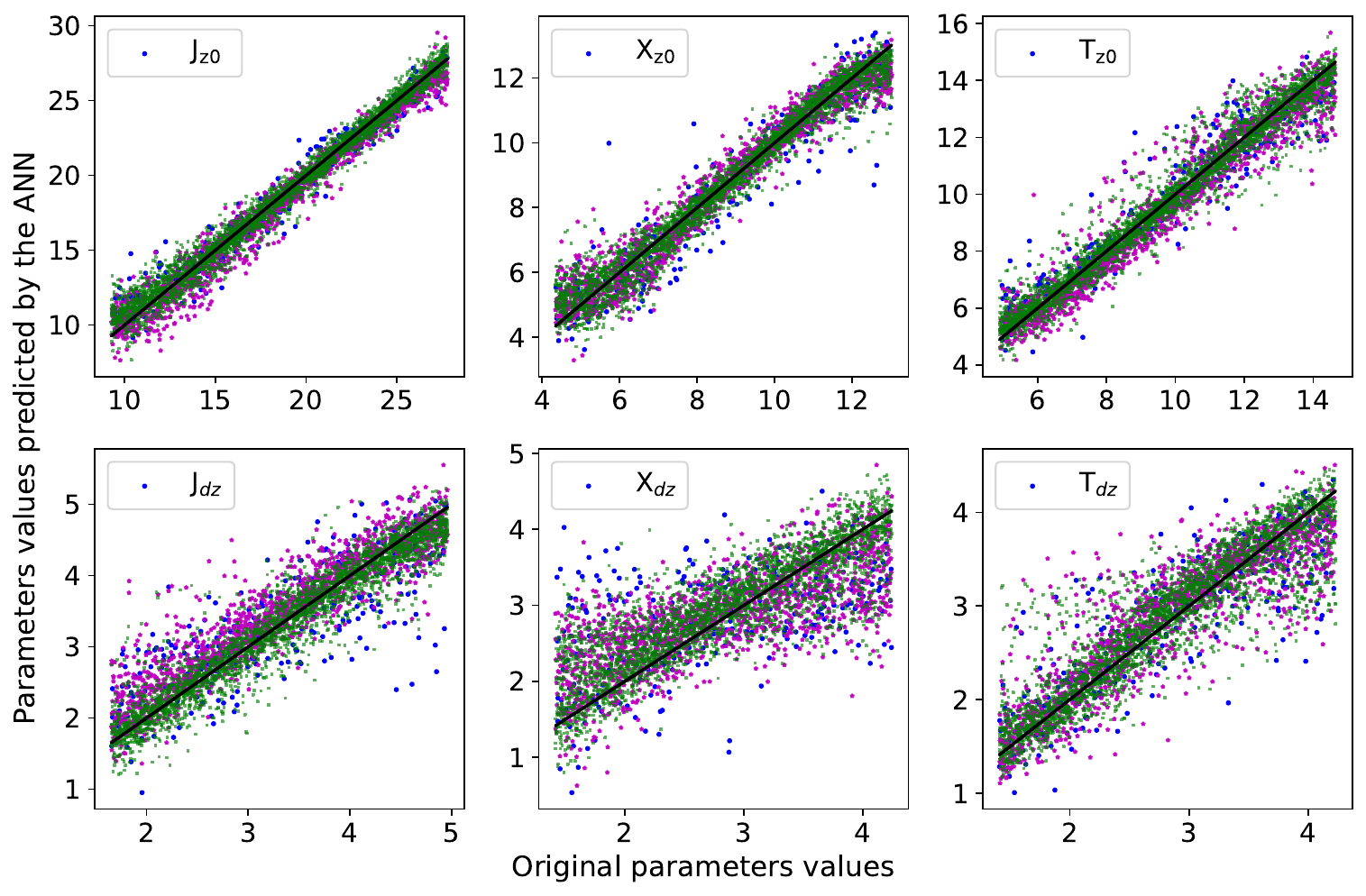}}
    \caption{The scatter plots above show predicted signal parameter values obtained through an ANN model trained on the global 21cm signal. The signal data sets are generated by sampling the parameter space using Hammersley sequence sampling in three sizes: 1000, 5000, and 10,000 samples. Blue points in each scatter plot denote predictions made by the ANN trained with 1000 samples, while magenta and green points indicate predictions from ANN models trained with 5000 and 10000 samples, respectively. The true value of the parameters are plotted in solid black line in the each plot.}
   
    \label{Fig3}
\end{figure*}

\begin{figure*}
    \centering
    {\includegraphics[width=1.0\textwidth]{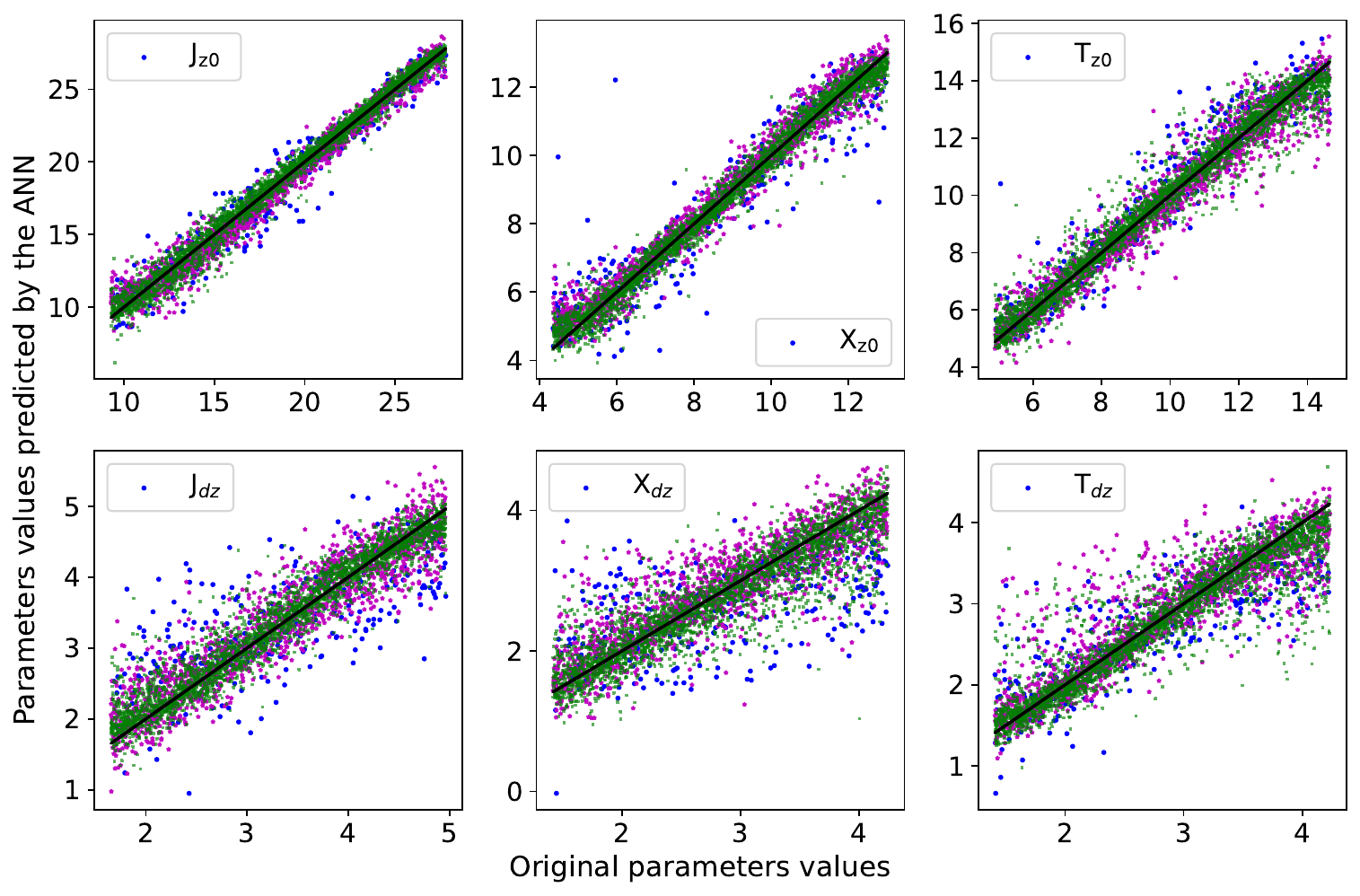}}
    \caption{The scatter plots above show predicted signal parameter values obtained through an ANN model trained on the parametrized global 21cm signal. The signal datasets are generated by sampling the parameter space using Latin hypercube sampling in three sizes: 1000, 5000, and 10,000 samples. Blue points in each scatter plot denote predictions made by the ANN trained with 1000 samples, while magenta and green points indicate predictions from ANN models trained with 5000 and 10000 samples, respectively. The true values of the parameters are plotted in a solid black line in each plot.}
   
    \label{Fig4}
\end{figure*}

\begin{figure*}
    \centering
    {\includegraphics[width=1.0\textwidth]{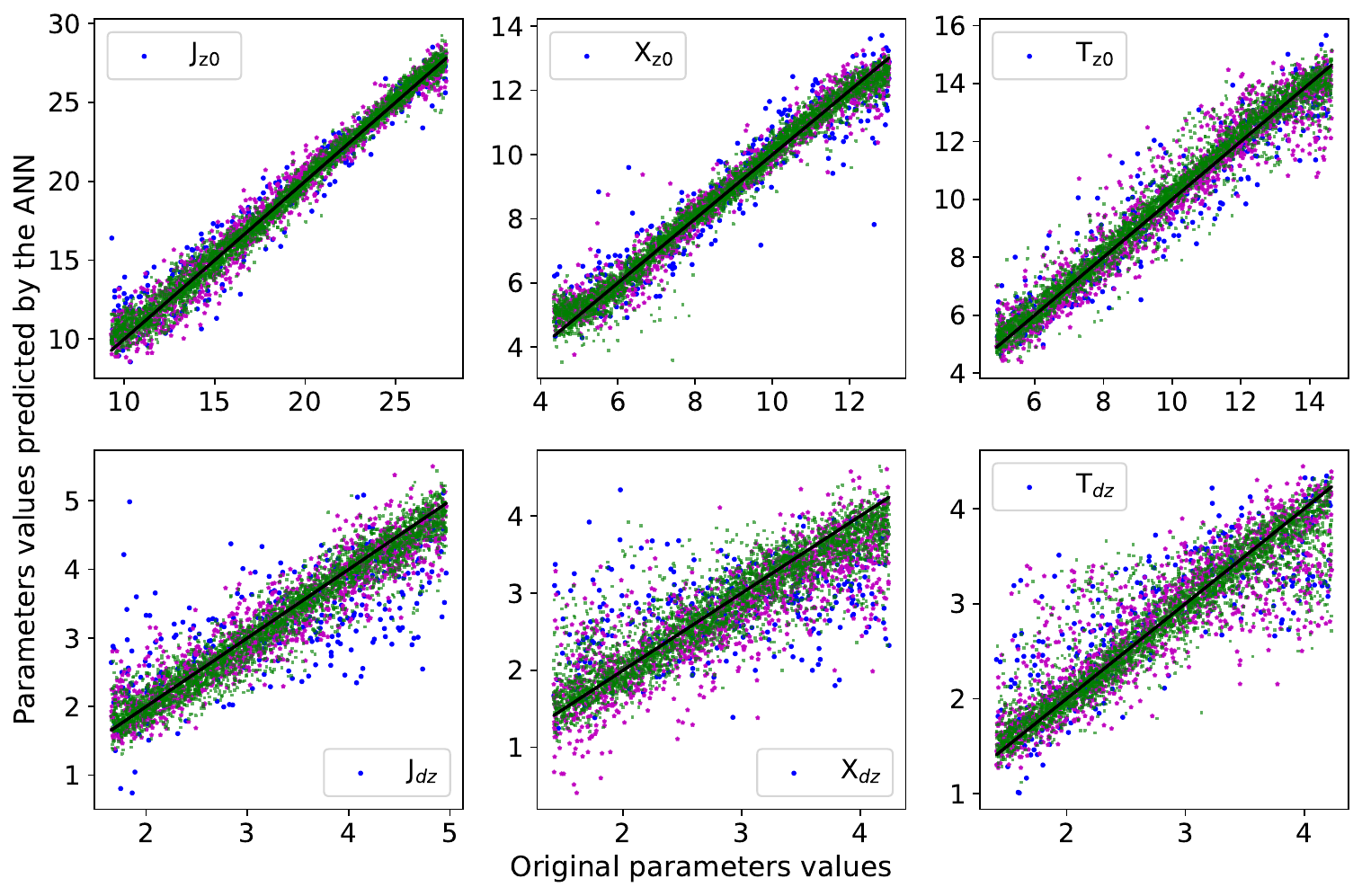}}
    \caption{The scatter plots above show predicted signal parameter values obtained through an ANN model trained on parametrized the global 21cm signal. The signal data sets are generated by sampling the parameter space using Random sampling in three sizes: 1000, 5000, and 10,000 samples. Blue points in each scatter plot denote predictions made by the ANN trained with 1000 samples, while magenta and green points indicate predictions from ANN models trained with 5000 and 10000 samples, respectively. The true values of the parameters are plotted in a solid black line in each plot.}
   
    \label{Fig5}
\end{figure*}

The ANN trained with 1000 datasets achieved the overall $\rm R^{2}$ score of 0.6744 for datasets sampled using HSS. For LHS, the overall $\rm R^{2}$ score obtained is 0.6594, while for Random sampling, the $\rm R^{2}$ score is obtained at 0.6367. Similarly, with 5000 datasets, the overall $\rm R^{2}$ score of 0.9059 was achieved for datasets sampled using HSS. For LHS, the overall $\rm R^{2}$ score obtained was 0.8937, while the overall $\rm R^{2}$ score of 0.8837 was obtained for the dataset sampled using Random sampling. With 10,000 datasets, the overall $\rm R^{2}$ score of 0.9259 was obtained for datasets sampled using HSS. For Random sampling, the overall $\rm R^{2}$ score obtained was 0.9187, while the overall $\rm R^{2}$ score of 0.9210 was obtained for the dataset sampled using LHS. The detailed results for each sampling method with the various dataset sizes, the $\rm R^2$ and RMSE score for the individual parameters are listed in Tab. \ref{tab1} and Tab. \ref{tab2}. We also individually show the predicted parameter values against the original values for each sampling method across different dataset sizes. These visualizations are presented in Fig. \ref{Fig3} for HSS, Fig. \ref{Fig4} for LHS, and Fig. \ref{Fig5} for Random sampling.

Our investigation revealed that, across all three sampling techniques, the accuracy of the ANN improved with an increase in the number of training datasets. For example, when dealing with datasets consisting of 10,000 samples, all three sampling methods demonstrated significantly higher prediction accuracy compared to situations where the network was trained with 5,000 and 1,000 samples using the same techniques. Further details are provided in Tab. \ref{tab1}. Additionally, it was observed that ANN models trained with Hammersley sequence-sampled datasets achieved slightly higher overall $\rm R^{2}$ scores for dataset sizes of 1000, 5000, and 10,000 compared to models trained with LHS and Randomly sampled datasets.

%%%%%%%%%%%%%%%%%%%%%%%%%%%%%%%%%%%%%%%%%%%%%%%%%%

\begin{table*}
\centering
\begin{tabular}{|l|l|l|l|l|l|l|l|l|l|}
\hline
Size   &          & 1000     &          &        &5000    &        &        &10000   &  \\ \hline
& HSS      & LHS      & Rand   & HSS    & LHS    & Rand & HSS    & LHS    & Rand  \\ \hline
Avg. & 0.6744   & 0.6594   & 0.6379  &0.9059 & 0.8937 & 0.8837 &0.9259 & 0.9187 & 0.9210  \\\hline
$\rm J_{z0 }$  & 0.9471   &  0.9442  & 0.9389   & 0.9821 & 0.9780 & 0.9770 & 0.9868 & 0.9844 & 0.9859  \\
$\rm X_{z0}$   & 0.8797   &  0.8552  & 0.8687   & 0.9716 & 0.9681 & 0.9693 & 0.9751 & 0.9793 & 0.9758  \\
$\rm T_{z0}$   & 0.8589   &  0.8899  & 0.8908   & 0.9449 & 0.9431 & 0.9410 & 0.9686 & 0.9583 & 0.9615  \\
$\rm J_{dz}$   & 0.5514   &  0.4943  & 0.4107   & 0.9094 & 0.8790 & 0.8978 & 0.9323 & 0.9294 & 0.9258  \\
$\rm X_{dz}$   & 0.1942   &  0.1372  & 0.0807   & 0.7982 & 0.7759 & 0.7200 & 0.8395 & 0.8075 & 0.8161  \\
$\rm T_{dz}$   & 0.6151   &  0.6377  & 0.6377   & 0.8289 & 0.8180 & 0.7973 & 0.8450 & 0.8534 & 0.8607  \\ \hline
\end{tabular}

\caption{The computed $\rm R^2$-scores for all signal parameters for predicted each case studied are listed here. We used the parametrized model to construct the global 21cm signal.}
\label{tab1}
\end{table*}

\begin{table*}
\centering
\begin{tabular}{|l|l|l|l|l|l|l|l|l|l|}
\hline
Size   &          & 1000     &          &        &5000    &        &        &10000   &  \\ \hline
& HSS      & LHS      & Rand   & HSS    & LHS    & Rand & HSS     & LHS    & Rand  \\ \hline
$\rm J_{z0 }$  & 0.0661   &  0.0679  & 0.0720   & 0.0392 & 0.0429 & 0.0447 & 0.0354  & 0.0356 & 0.0346  \\
$\rm X_{z0}$   & 0.1044   &  0.1137  & 0.1035   & 0.0502 & 0.0513 & 0.0508 & 0.0460  & 0.0420 & 0.0452  \\
$\rm T_{z0}$   & 0.1077   &  0.0944  & 0.0980   & 0.0680 & 0.0678 & 0.0716 & 0.0586  & 0.0588 & 0.0572  \\
$\rm J_{dz}$   & 0.1905   &  0.1997  & 0.2211   & 0.0876 & 0.1018 & 0.0920 & 0.0754  & 0.0772 & 0.0788 \\
$\rm X_{dz}$   & 0.2661   &  0.2740  & 0.2657   & 0.1303 & 0.1347 & 0.1527 & 0.1153  & 0.1265 & 0.1238  \\
$\rm T_{dz}$   & 0.1805   &  0.1710  & 0.1754   & 0.1186 & 0.1242 & 0.1294 & 0.1118  & 0.1118 & 0.1060  \\ \hline
\end{tabular}

\caption{The computed RMSE scores for all signal parameters for predicted each case studied are listed here. We used the parametrized model to construct the global 21cm signal.}
\label{tab2}
\end{table*}

\subsection{Signal and foreground with thermal noise}
In this study, we trained an ANN using datasets that included signals corrupted with foreground and thermal noise. The addition of foreground into the global 21cm signal not only increases the complexity of the dataset but also expands the dimensionality of the parameter space. We followed a methodology similar to that used for signal data to construct our training datasets. 
We charted the parameter space for both the signal and foreground components using three different sampling methods. We generated three distinct dataset sizes for the parametrized signal: 10,000, 50,000, and 200,000 samples to explore the minimum sample size required to train the ANN for improved accuracy and robustness effectively. We pursued two distinct approaches to train the ANN with a signal corrupted by foreground and thermal noise. In the first method, we normalized and standardized with 'MinMaxScaler' and `StandardScaler'; in the second method, we logarithmically scaled the training dataset. Subsequently, we normalized and standardized the datasets further. In the first case, we achieved better accuracy in recovering only two foreground parameters, $a_{1}$ and $a_{2}$, while the rest of the parameters are recovered poorly. In the second case, we successfully recovered all parameters with reasonable accuracy, except for the two foreground parameters $a_{1}$ and $a_{2}$, where the accuracy ranged from 50 \% to 60 \% in the best-case scenario. Based on these experiences, we decided to train two separate ANN models for both parametrized and physical cases. The first ANN model focused on the recovery of the two foreground parameters ($a_{1}, a_{2}$) without the use of logarithmically scaled datasets. For the remaining parameters, the second ANN model utilized logarithmically scaled datasets. A detailed description of the architectures of these ANN models is described below.

The first ANN model's architecture featured an input layer with 1024 neurons, aligning with the data's 1024 frequency channels. This was followed by two hidden layers with 64 and 16 neurons, respectively, and the output layer with 2 neurons to predict the two foreground parameters ($a_{1}$, $a_{2}$). In the case of the second ANN model, the input layer mirrored the first model with 1024 neurons. Following this, four hidden layers were introduced, consisting of 256, 64, 32, and 16 neurons. The output layer of the second ANN model comprised 8 neurons, each representing 6 signal parameters and 2 foreground parameters ($a_{0}$, $a_{3}$). In both models, each hidden layer utilized the Exponential Linear Unit (ELU) activation function. To mitigate overfitting, both ANN models employed the `normal' kernel initializer and implemented L2 kernel regularization. Before training the ANN models in both scenarios, we normalized the parameters using the `MinMaxScaler' scaling.

\begin{figure*}
    \centering
    {\includegraphics[width=1.0\textwidth]{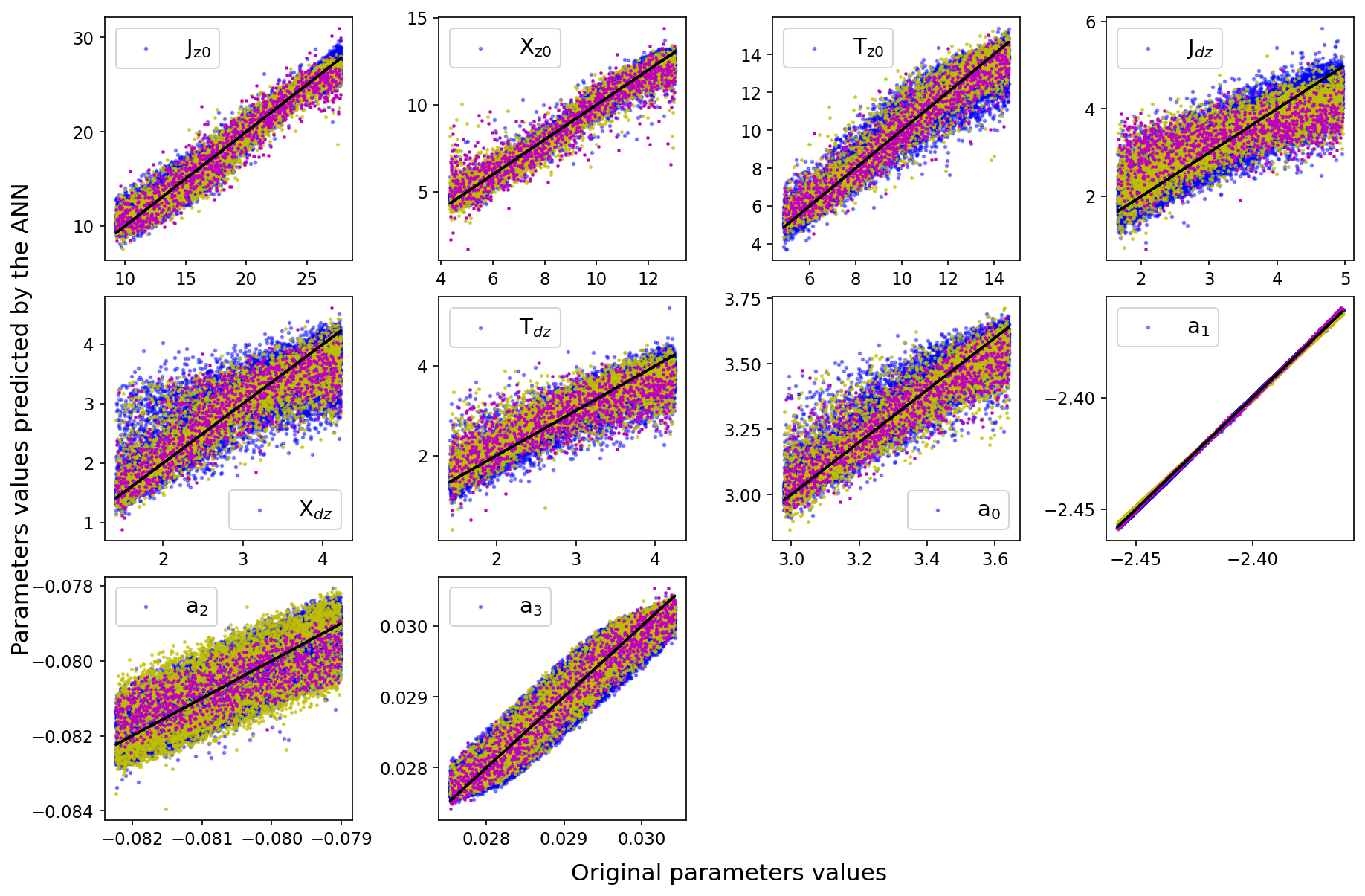}}
    
    \caption{The scatter plots above show predicted signal and foreground parameter values obtained through an ANN model trained on a parametrized global 21cm signal. The signal and foreground data sets are generated by sampling the parameter space using Hammersley sequence sampling in three sizes: 10,000, 50,000, and 200,000 samples. Magenta scatter points in each scatter plot denote predictions made by the ANN trained with 10,000 samples, while yellow and blue scatter points indicate predictions from ANN models trained with 50,000 and 200,000 samples, respectively. The actual values of the parameters are plotted in a solid black line in each plot.}
   
    \label{Fig9}
\end{figure*}

\begin{figure*}
    \centering
    {\includegraphics[width=1.0\textwidth]{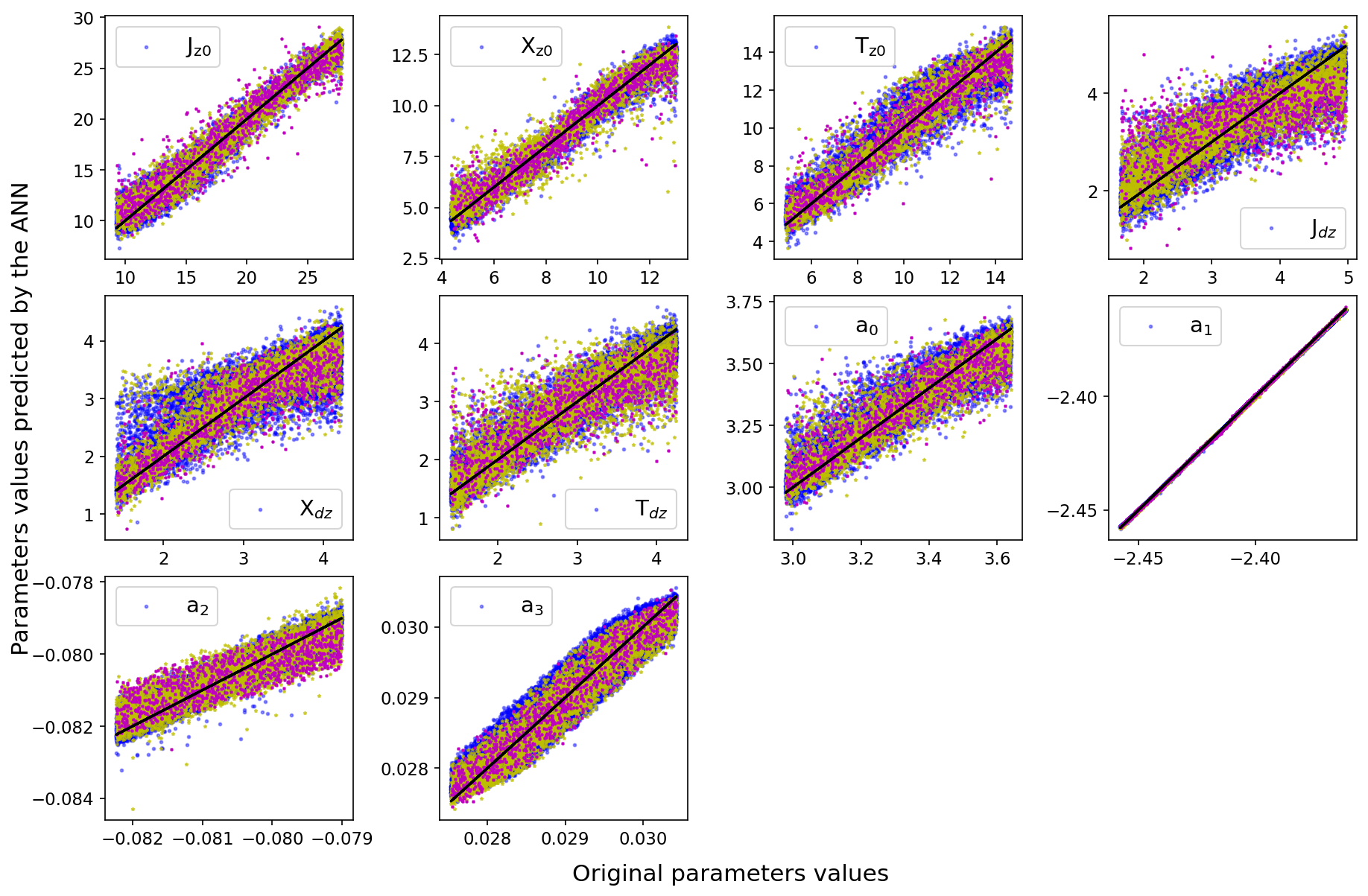}}
    
    \caption{The scatter plots above show predicted signal and foreground parameter values obtained through an ANN model trained on a parametrized global 21cm signal. The signal and foreground data sets are generated by sampling the parameter space using Latin hypercube sampling in three sizes: 10,000, 50,000, and 200,000 samples. Magenta scatter points in each scatter plot denote predictions made by the ANN trained with 10,000 samples, while yellow and blue scatter points indicate predictions from ANN models trained with 50,000 and 200,000 samples, respectively. The actual values of the parameters are plotted in a solid black line in each plot.}
   
    \label{Fig10}
\end{figure*}

\begin{figure*}
    \centering
    {\includegraphics[width=1.0\textwidth]{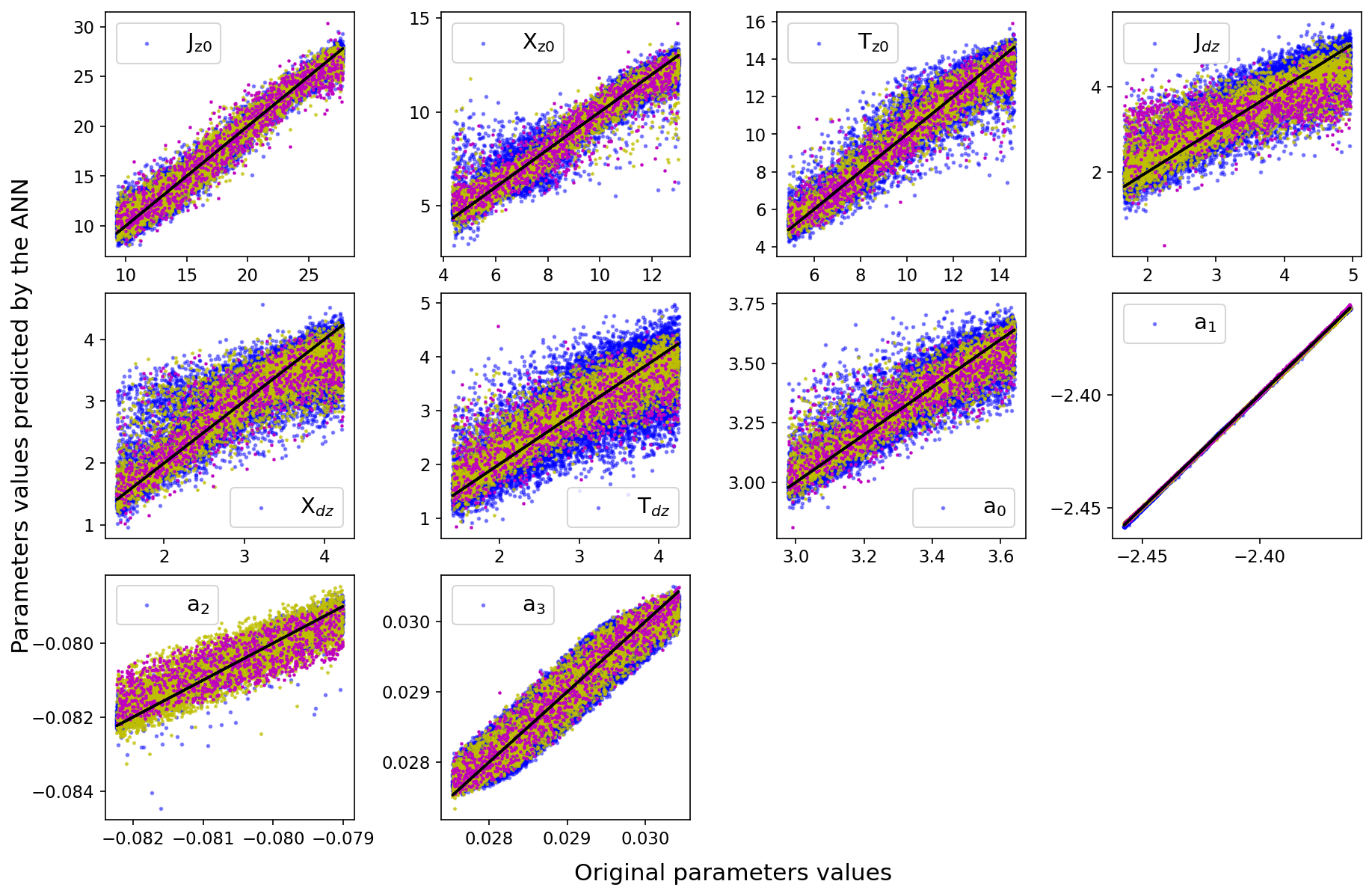}}
    
    \caption{The scatter plots above show predicted signal and foreground parameter values obtained through an ANN model trained on a parametrized global 21cm signal. The signal and foreground data sets are generated by sampling the parameter space using Random sampling in three sizes: 10,000, 50,000, and 200,000 samples. Magenta scatter points in each scatter plot denote predictions made by the ANN trained with 10,000 samples, while yellow and blue scatter points indicate predictions from ANN models trained with 50,000 and 200,000 samples, respectively. The actual values of the parameters are plotted in a solid black line in each plot.}
   
    \label{Fig11}
\end{figure*}

The ANN trained with 10,000 datasets achieved the overall $\rm R^{2}$ score of 0.7534 for datasets sampled using LHS. For Random, the overall $\rm R^{2}$ score obtained was 0.7307, while the $\rm R^{2}$ score of 0.7307 was obtained for the dataset sampled using HSS. Similarly, with 50,000 datasets, the overall $\rm R^{2}$ score of 0.8738 was achieved for datasets sampled using Random sampling. For LHS, the overall $\rm R^{2}$ score obtained was 0.8679, while the overall $\rm R^{2}$ score of 0.8673 was obtained for the sampled dataset using HSS. With 200,000 datasets, the overall $\rm R^{2}$ score of 0.9296 was attained for datasets sampled using LHS. For HSS, the overall $\rm R^{2}$ score obtained was 0.9139, while the $\rm R^{2}$ score of 0.9016 was obtained for the sampled dataset using Random sampling. The detailed results for each sampling method with the various dataset sizes, the $ \rm R^2$ and RMSE score for individual parameters are listed in Tab. \ref{tab9} and Tab. \ref{tab10}. We also individually showed the predicted parameter values against the original values for each sampling method across different dataset sizes. These visualizations are presented in Fig. \ref{Fig9} for HSS, Figure \ref{Fig10} for LHS, and Fig. \ref{Fig11} for Random sampling. 

We noticed that as we increased the complexity and dimensionality of the problem, achieving an optimal solution with the ANN required drawing more sample sets to cover the entire parameter space. In contrast to the scenario where only the signal was considered, where optimal ANN prediction was achieved with 10,000 datasets, here, with the same number of datasets, we only achieved an accuracy of around $\sim 75\%$, regardless of the sampling method. For this particular case, to attain a similar level of accuracy in prediction, we found that drawing approximately 200,000 samples was necessary.

\begin{table}
\centering
\begin{tabular}{|l|l|l|l|l|l|l|l|l|l|}
\hline
Size   &          & 10000         &          &        &50000    &        &        &200000   &  \\ \hline
     & HSS      & LHS          & Rand   & HSS    & LHS                  & Rand  & HSS    & LHS            & Rand  \\ \hline
Total & 0.7252  & 0.7534 & 0.7307           & 0.8673 & 0.8679    & 0.8738  & 0.9139  & 0.9296 & 0.9016  \\ \hline
$\rm J_{z0 }$  & 0.8993  & 0.9020   & 0.9109   & 0.9573 & 0.9659    & 0.9626   & 0.9753      & 0.9798          &0.9784  \\
$\rm X_{z0}$   & 0.8297  & 0.8962   & 0.8414   & 0.9447 & 0.9408    & 0.9447   & 0.9798      & 0.9796          & 0.9461  \\
$\rm T_{z0}$   & 0.8875  & 0.8689   & 0.8634   & 0.9315 & 0.9336    & 0.9396   & 0.9483      & 0.9554          & 0.9434  \\
$\rm J_{dz}$   & 0.2186  & 0.2677   & 0.2813   & 0.6651 & 0.7356    & 0.7244   & 0.8246      & 0.8603          & 0.8328  \\
$\rm X_{dz}$   & 0.6568  & 0.6968   & 0.6595   & 0.7838 & 0.7934    & 0.8025   & 0.8248      & 0.8458          & 0.8361  \\
$\rm T_{dz}$   & 0.4979  & 0.5837   & 0.5005   & 0.7953 & 0.7472    & 0.7875   & 0.8896      & 0.9083          & 0.7240  \\ 
$\rm a_{0}$    & 0.7229  & 0.7452   & 0.7292   & 0.8375 & 0.8672    & 0.8773   & 0.9086      & 0.9214          & 0.8889  \\
$\rm a_{1}$    & 0.9994  & 0.9997   & 0.9996   & 0.9989 & 0.9997    & 0.9999   & 0.9994      & 0.9999          & 0.9994  \\
$\rm a_{2}$    & 0.6095  & 0.6460   & 0.5964   & 0.8274 & 0.7659    & 0.7684   & 0.8606      & 0.9152          &0.9363   \\
$\rm a_{3}$    & 0.9298  & 0.9281   & 0.9248   & 0.9319 & 0.9296    & 0.9311   & 0.9277      & 0.9302          &0.9304    \\ \hline
\end{tabular}

\caption{The computed $\rm R^2$-scores for all signal and foreground parameters for each case studied are listed here. We used the parametrized model to construct the global 21cm signal and the log-log polynomial to construct the foreground.}
\label{tab9}
\end{table}

\begin{table*}
\centering
\begin{tabular}{|l|l|l|l|l|l|l|l|l|l|}
\hline
Size   &          & 10000     &          &        &50000    &        &        &200000   &  \\ \hline
                 & HSS      & LHS      & Rand     & HSS    & LHS    & Rand    & HSS    & LHS    & Rand  \\ \hline
$\rm J_{z0 }$  & 0.0917   &  0.0914  & 0.0861   & 0.0593 & 0.0532 & 0.0556   & 0.0456 & 0.0408 & 0.0423  \\
$\rm X_{z0}$   & 0.1180   &  0.0904  & 0.1139   & 0.0679 & 0.0709 & 0.0673   & 0.0409 & 0.0412 & 0.0670  \\
$\rm T_{z0}$   & 0.0980   &  0.1049  & 0.1063   & 0.0750 & 0.0737 & 0.0705   & 0.0659 & 0.0610 & 0.0683  \\
$\rm J_{dz}$   & 0.2567   &  0.2426  & 0.2442   & 0.1655 & 0.1494 & 0.1513   & 0.1211 & 0.1082 & 0.1185  \\
$\rm X_{dz}$   & 0.1651   &  0.1580  & 0.1666   & 0.1332 & 0.1307 & 0.1281   & 0.1210 & 0.1133 & 0.1166  \\
$\rm T_{dz}$   & 0.2019   &  0.1828  & 0.2016   & 0.1310 & 0.1453 & 0.1336   & 0.0961 & 0.0874 & 0.1508  \\ 
$\rm a_{0}$    & 0.1480   &  0.1464  & 0.1500   & 0.1170 & 0.1051 & 0.1008   & 0.0870 & 0.0808 & 0.0964\\
$\rm a_{1}$    & 0.0068   &  0.0048  & 0.0053   & 0.0092 & 0.0043 & 0.0028   & 0.0066 & 0.0021 & 0.0065\\
$\rm a_{2}$    & 0.1803   &  0.1750  & 0.1833   & 0.1189 & 0.1400 & 0.1394   & 0.1073 & 0.0840 & 0.0728\\
$\rm a_{3}$    & 0.0758   &  0.0769  & 0.0778   & 0.0750 & 0.0762 & 0.0759   & 0.0077 & 0.0760 & 0.0759\\ \hline
\end{tabular}

\caption{The computed $\rm RMSE $-scores for all signal and foreground parameters for each case studied are listed here. We used the parametrized model to construct the global 21cm signal and the log-log polynomial to construct the foreground.}
\label{tab10}
\end{table*}

\subsection{Generalizability test of the ANN models}
To assess the robustness and generalizability of our trained ANN models, we conducted tests using separate unknown test datasets generated via three distinct sampling techniques. This comprehensive analysis aims to identify the most robustly trained ANN model among those trained with differently sampled datasets. Throughout the training process, we identified the likelihood of bias when drawing conclusions based on a single instance of training the ANN model. This recognition stemmed from factors including variations in initial sample seeds,  clustering issues, dataset partitioning, and the inherent variability in training the ANN itself. To mitigate this bias and ensure the consistency of our results across multiple trials, we repeated the entire training and sampling procedure 11 times and saved these individual ANN models for further testing with unknown test datasets drawn via these three distinct sampling techniques: HSS, LHS and Random. 

\begin{figure*}
    \centering
    {\includegraphics[width=0.9\textwidth]{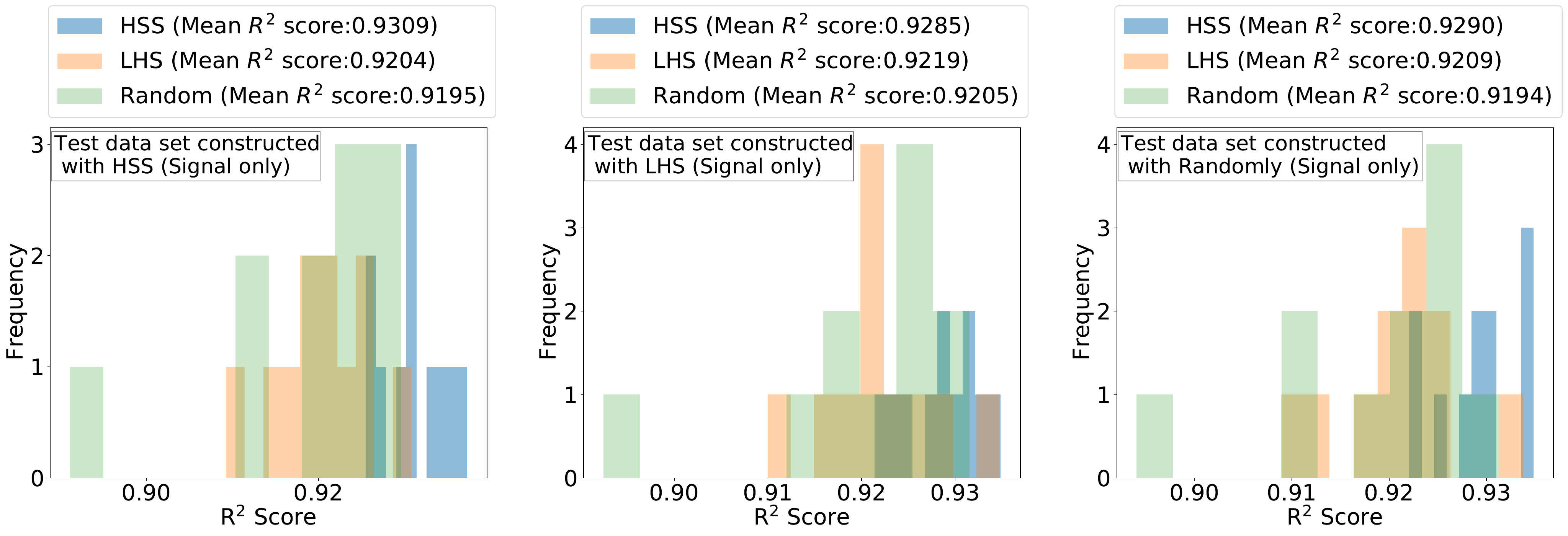}}
    {\includegraphics[width=0.9\textwidth]{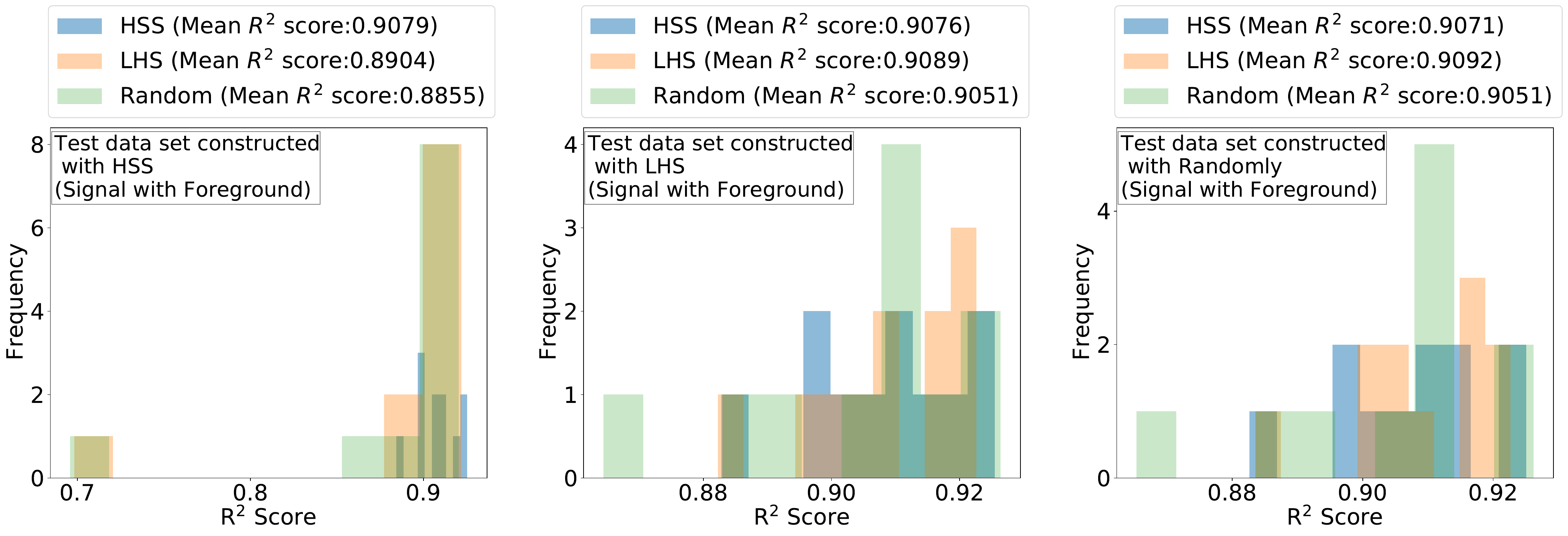}}
    %{\includegraphics[width=0.9\textwidth]{Image/Robustness/Physical_sig.pdf}}
    %{\includegraphics[width=0.9\textwidth]{Image/Robustness/Physical_sig_fore.pdf}}
    \caption{The figure illustrates ANN model predictions for various trials trained on datasets sampled using HSS, LHS, and Random methods with optimal sample sizes. Histograms depict ANN prediction accuracy measured by $\rm R^{2}$ scores. Panels in the figure's rows correspond to different signal scenarios: the top three panels in the first row show predictions for the Parameterized signal, followed by predictions for the Parameterized signal with foreground and thermal noise in the second row.  Histograms are colour-coded: blue represents HSS-trained ANN models, orange represents LHS-trained ANN models, and green represents the Random-trained ANN model's prediction accuracy. }
    \label{Fig_robustness}
\end{figure*}

We computed the mean prediction accuracy in terms of $\rm R^{2}$ and found that ANNs trained on Hammersley-sampled datasets exhibited lower fluctuations from the mean $\rm R^{2}$ value across test datasets, compared to those trained on LHS and Random datasets. For instance, the ANN trained on HSS-sampled datasets achieved a mean $\rm R^{2}$ score of 0.9309, with a range of 0.9255 to 0.9373. Similarly, for LHS-sampled test datasets, the mean $\rm R^{2}$ score was 0.9285 (range: 0.9213 to 0.9348), and for Random-sampled test datasets, it was 0.9290 (range: 0.9221 to 0.9348). In contrast, ANNs trained on LHS-sampled datasets had a mean $\rm R^{2}$ score of 0.9204 on HSS-sampled test datasets (range: 0.9093 to 0.9308), 0.9219 on LHS-sampled test datasets (range: 0.9100 to 0.9347), and 0.9209 on Random-sampled test datasets (range: 0.9089 to 0.9347). For ANNs trained on Random-sampled datasets, the mean $\rm R^{2}$ scores were 0.9195 on HSS-sampled test datasets (range: 0.8911 to 0.9296), 0.9205 on LHS-sampled test datasets (range: 0.8924 to 0.9315), and 0.9194 on Random-sampled test datasets (range: 0.8940 to 0.9313). For a detailed visual representation of these results, refer to Fig. \ref{Fig_robustness}, depicted in the first row. The histogram plots are color-coded: blue represents the prediction accuracy, in terms of $\rm R^{2}$ score, of HSS-trained ANN models; orange represents ANN models trained with LHS-sampled datasets, and green represents the $\rm R^{2}$ score histogram for ANN models trained with Random-sampled datasets.

Similarly, for the parametrized  signal and foreground case, ANN models trained with Hammersley sampled datasets consistently exhibited high prediction accuracy across all test datasets, with the highest $\rm R^{2}$ score approximately around 0.92 and the lowest around 0.88, with a mean of approximately 0.91 for each case. Detailed results are depicted in Fig. \ref{Fig_robustness} in the second row. Conversely, ANN models trained with Latin hypercube and Random sampled datasets showed inconsistent prediction accuracy for both cases. For test datasets generated with HSS, the lowest $\rm R^{2}$ score was less than 0.70, approximately 19-20 \% lower than the mean $\rm R^{2}$ score, approximately 0.90. Similar fluctuations were observed for test datasets constructed using Latin hypercube and Random sampling methods; detailed results are presented in Fig. \ref{Fig_robustness}, with the colour scheme remaining consistent across all subplots. Similarly, for the non-parametrized  signal case, we observed consistent behaviour, with ANN models trained on Hammersley sampled datasets exhibiting less deviation from the mean $\rm R^{2}$ score compared to models trained on LHS and Random datasets, the detailed result present in Appendix \ref{appendix_generalizability}.

\section{Summary and discussions}\label{summary_discussion}
In this study, our objective was to systematically explore the vast parameter space encompassing the 21cm global signal and foreground. Our goal was to simulate all possible variations using these parameters and then utilize the generated datasets to enhance the resilience of our trained ANN model. Recognizing the computational challenges posed by considering every possible parameter combination, we tested diverse sampling techniques to map the parameter space efficiently. Determining the ideal number of training datasets necessary for robust ANN model training is not straightforward due to the complexity of the model and the multitude of free parameters involved. Therefore, our research delved into a detailed analysis to establish the minimum dataset requirement. This exploration was essential in understanding how the number of datasets correlates with the model's complexity and the number of free parameters, ensuring comprehensive coverage of the parameter space and enabling the robust training of the ANN model. In our study, we explored various aspects by mapping the parameter space using three distinct sampling methods: Random, Latin hypercube, and Hammersley sequence sampling. Additionally, we investigated different sample sizes within specified boundary conditions to determine the minimum number of samples required to train the ANN effectively. To evaluate the performance of these sampling methods, we calculated metrics such as the $\rm R^{2}$ score and Root Mean Square Error (RMSE) using the ANN model. Additionally, we test the ANN's performance across different dataset sizes generated through these sampling techniques. These are some key findings noted from this study:

\begin{itemize}
    \item Regardless of the sampling method, the ANN's performance improved when trained with larger, well-sampled datasets. For instance, in parametrized signal scenarios, training with 1000 samples yielded an $\rm R^{2}$ score of 0.6744, while 5000 samples improved the score to 0.9059. Increasing the sample size to 10,000 resulted in a marginal improvement to 0.9226, indicating diminishing returns relative to the computational cost.
    \item The number of free parameters played a critical role; fewer parameters required fewer samples for optimal results, while more parameters necessitated larger sample sizes. For example, in the case of parametric signals with 6 free parameters, 10,000 datasets were sufficient for optimal results. However, for foreground-corrupted signals with 10 free parameters, 200,000 datasets were needed to achieve the same level of accuracy. This trend persisted when introducing different signal models as well.
    \item Models trained with datasets from HSS showed consistent performance across various unknown test datasets, regardless of the sampling method used. Conversely, models trained with LHS and Random methods exhibited inconsistent prediction accuracy, indicating less robustness.
    \item In lower dimensions ($< 10$), the ANN trained with HSS sampled datasets demonstrated slightly higher accuracy than those trained with LHS and Random methods. However, in higher-dimensional parameter spaces (  $\ge 10$), HSS performance declined due to clustering issues. For instance, when dealing with signal, foreground, and noise in parametric signal cases, HSS performance was slightly lower than the other sampling methods.
    \item We also obtained consistent results when using a non-parametric signal model instead of a parametric one. The primary difference observed was that, since the non-parametric model only had three free parameters for signal modeling, optimal ANN performance was achieved with a smaller dataset size. This finding suggests that the efficacy of sampling methods is influenced more by the dimensionality of the parameter space associated with the signal rather than by the specific signal model employed.
\end{itemize}

To address the high-dimensionality limitations of Hammersley sequence sampling, future research will explore alternative sampling methods that may offer viable solutions. Additionally, we plan to consider the ionospheric effect and beam chromaticity effect, investigating how these parameters impact the parameter space when combined with signal and foreground parameters. This investigation will help us to determine the minimum number of samples required for robust training of the ANN model, enabling accurate parameter inference from real observational datasets. In this study, we have compared these sampling methods solely within an ANN-based framework. In future research, we plan to incorporate other machine learning regression models to evaluate and compare their performance.

\section*{Acknowledgements}
AT would like to thank the Indian Institute of Technology Indore for providing funding for this study in the form of a Teaching Assistantship. SM and AD acknowledge financial support through the project titled ``Observing the Cosmic Dawn in Multicolour using Next Generation Telescopes'' funded by the Science and Engineering Research Board (SERB), Department of Science and Technology, Government of India through the Core Research Grant No. CRG/2021/004025. The authors acknowledge the use of facilities procured through the funding via the Department of Science and Technology, Government of India sponsored DST-FIST grant no. SR/FST/PSII/2021/162 (C) awarded to the DAASE, IIT Indore.

\appendix
\section{Non parametric model} \label{Non_parametric}
We have conducted similar analyses across different signal model to demonstrate their generalizability. The comprehensive results are discussed below.
\subsection{Signal only}
To ensure consistency and generalizability of the sampling techniques, we employ a different signal model, a non-parametrized signal model. This allows us to highlight any biases of the sampling techniques over the signal model, and assess performance in lower dimensions due to fewer free parameters. In contrast to the parametric model, the architecture of the ANN for this case is different. The input layer consisted of 1024 neurons, corresponding to the 1024 frequency channels. There were two hidden layers with 64 and 16 neurons, respectively, and these layers were activated using the 'sigmoid' and 'relu' activation functions. To prevent overfitting, we also applied L2 kernel regularization. The output layer has 3 neurons, each representing different signal parameters. We have used Adam optimizer with a learning rate of $10^{-4}$. 

\begin{figure*}
    \centering
    {\includegraphics[width=1.0\textwidth]{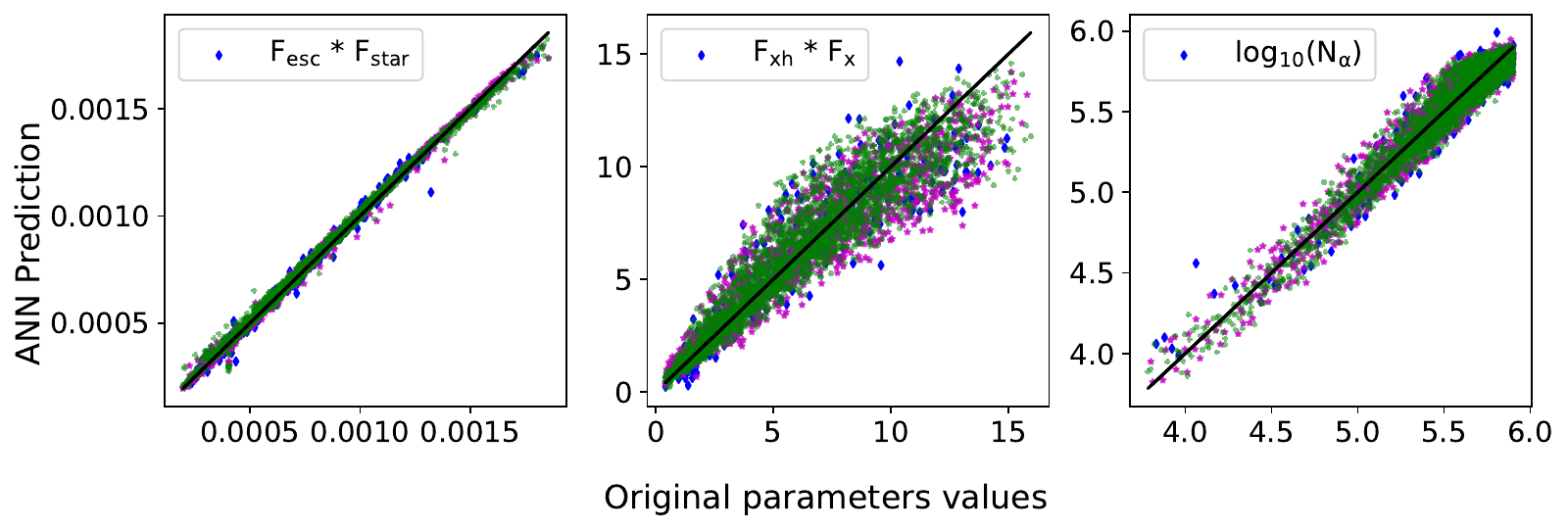}}
    \caption{The scatter plots above show predicted signal parameter values obtained through an ANN model trained on a physical global 21cm signal. The signal data sets are generated by sampling the parameter space using Hammersley sequence sampling in three sizes: 1000, 5000, and 10,000 samples. Blue points in each scatter plot denote predictions made by the ANN trained with 1000 samples, while magenta and green points indicate predictions from ANN models trained with 5000 and 10000 samples, respectively. The actual value of the parameters is plotted in a solid black line in each plot.}
   
    \label{Fig6}
\end{figure*}

\begin{figure*}
    \centering
    {\includegraphics[width=1.0\textwidth]{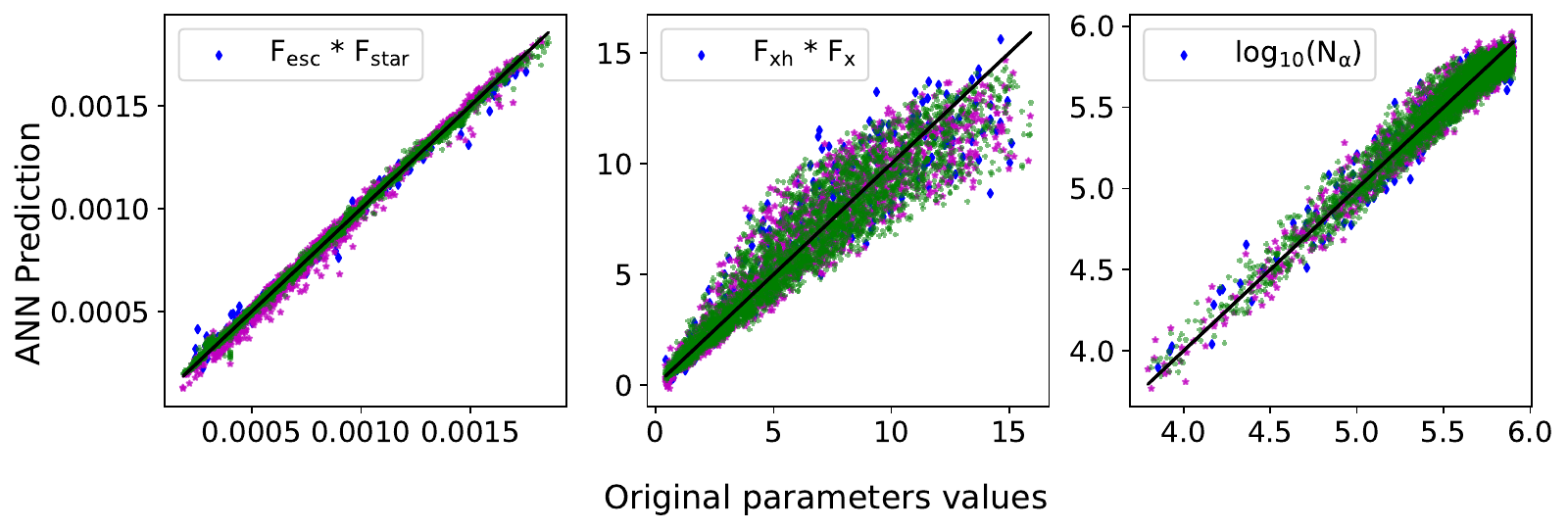}}
    \caption{The scatter plots above show predicted signal parameter values obtained through an ANN model trained on a physical global 21cm signal. The signal data sets are generated by sampling the parameter space using Latin hypercube sampling in three sizes: 1000, 5000, and 10,000 samples. Blue points in each scatter plot denote predictions made by the ANN trained with 1000 samples, while magenta and green points indicate predictions from ANN models trained with 5000 and 10000 samples, respectively. The actual values of the parameters are plotted in a solid black line in each plot.}
   
    \label{Fig7}
\end{figure*}

\begin{figure*}
    \centering
    {\includegraphics[width=1.0\textwidth]{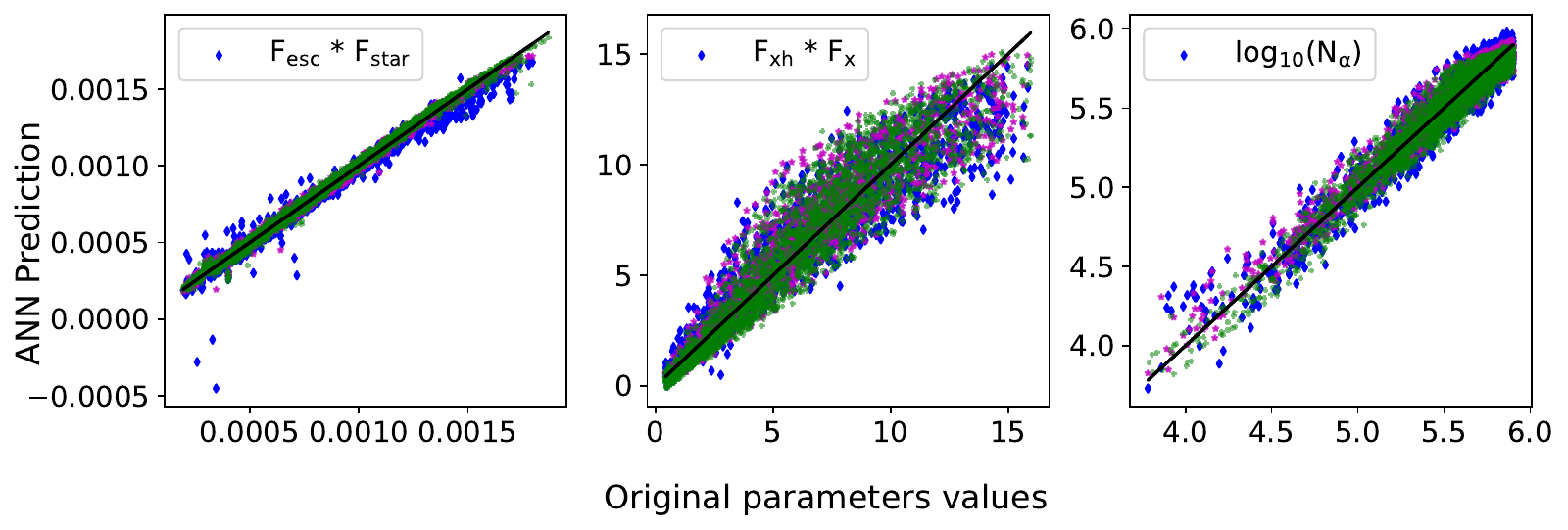}}
    \caption{The scatter plots above show predicted signal parameter values obtained through an ANN model trained on a physical global 21cm signal. The signal data sets are generated by sampling the parameter space using Random sampling in three sizes: 1000, 5000, and 10,000 samples. Blue points in each scatter plot denote predictions made by the ANN trained with 1000 samples, while magenta and green points indicate predictions from ANN models trained with 5000 and 10000 samples, respectively. The actual value of the parameters is plotted in a solid black line in each plot.}
   
    \label{Fig8}
\end{figure*}

Similar to the parametric context, a consistent trend is observed in the non-parametric signal scenario: the ANN model trained with 1000 datasets achieves an overall $\rm R^{2}$ score of around 0.92, which increases to approximately 0.93 with 5000 datasets, and around 0.94 with 10,000 datasets. Detailed results for each sampling method and dataset size, including $R^2$ and RMSE scores, are provided in Tab. \ref{tab3} and Tab. \ref{tab4}. Visualizations comparing predicted parameter values against original values for each sampling method are displayed in Fig. \ref{Fig6} (HSS), Fig. \ref{Fig7} (LHS), and Fig. \ref{Fig8} (Random sampling). Notably, model performance improves with larger sample sizes, with all methods achieving comparable accuracy levels for substantial datasets. The lower dimensionality of the problem compared to the parametric case allows for higher prediction accuracy with fewer sampled datasets.

\begin{table*}
\centering
\begin{tabular}{|l|l|l|l|l|l|l|l|l|l|}
\hline
Size       &        & 1000        &         &                 &5000    &                      &     &10000 &  \\ \hline
        & HSS     & LHS        & Rand   & HSS     & LHS    & Rand           & HSS             & LHS  & Rand   \\  \hline
Avg.        & 0.9155 & 0.9252 & 0.9214   & 0.9392  & 0.9367 & 0.9346  &0.9447 & 0.9446 & 0.9418 \\ \hline
$f_{\ast}.f_{esc}$  & 0.9936  & 0.9922     & 0.9893   & 0.9945  & 0.9957 & 0.9940           & 0.9974         & 0.9970 & 0.9944 \\
$f_{X,h}.f_{X}$     & 0.8255  & 0.8373     & 0.8408   & 0.8785  & 0.8682 & 0.8642           & 0.8837         & 0.8737 & 0.8783  \\
$N_{\alpha}$        & 0.9275  & 0.9454     & 0.9341   & 0.9531  & 0.9539 & 0.9555           & 0.9531         & 0.9532 & 0.9527 \\  \hline
\end{tabular}

\caption{The computed $\rm R^2$-scores for all signal parameters for each case studied are listed here. We used the physical model to construct the global 21cm signal.}
\label{tab3}
\end{table*}

\begin{table*}
\centering
\begin{tabular}{|l|l|l|l|l|l|l|l|l|l|}
\hline
Size        &        & 1000     &         &         &5000 &        &     &10000 &  \\ \hline
                      & HSS     & LHS     & Rand   & HSS     & LHS   & Rand  & HSS    & LHS  & Rand   \\  \hline
$f_{\ast}.f_{esc}$  & 0.0230  & 0.0253  & 0.0281   & 0.0149 & 0.0162 & 0.0162 & 0.0130 & 0.0135 & 0.0150 \\
$f_{X,h}.f_{X}$     & 0.1154  & 0.1000  & 0.1130   & 0.1022 & 0.1064 & 0.1043 & 0.0875 & 0.0851 & 0.0800  \\
$N_{\alpha}$        & 0.0554  & 0.0482  & 0.0530   & 0.0447 & 0.0421 & 0.0421 & 0.0408 & 0.0392 & 0.0390 \\  \hline
\end{tabular}
\caption{The computed RMSE scores for all signal parameters for each case studied are listed here. We used the physical model to construct the global 21cm signal.}
\label{tab4}
\end{table*}

\subsection{Signal with foreground and thermal noise}
\begin{figure*}
    \centering
    {\includegraphics[width=1.0\textwidth]{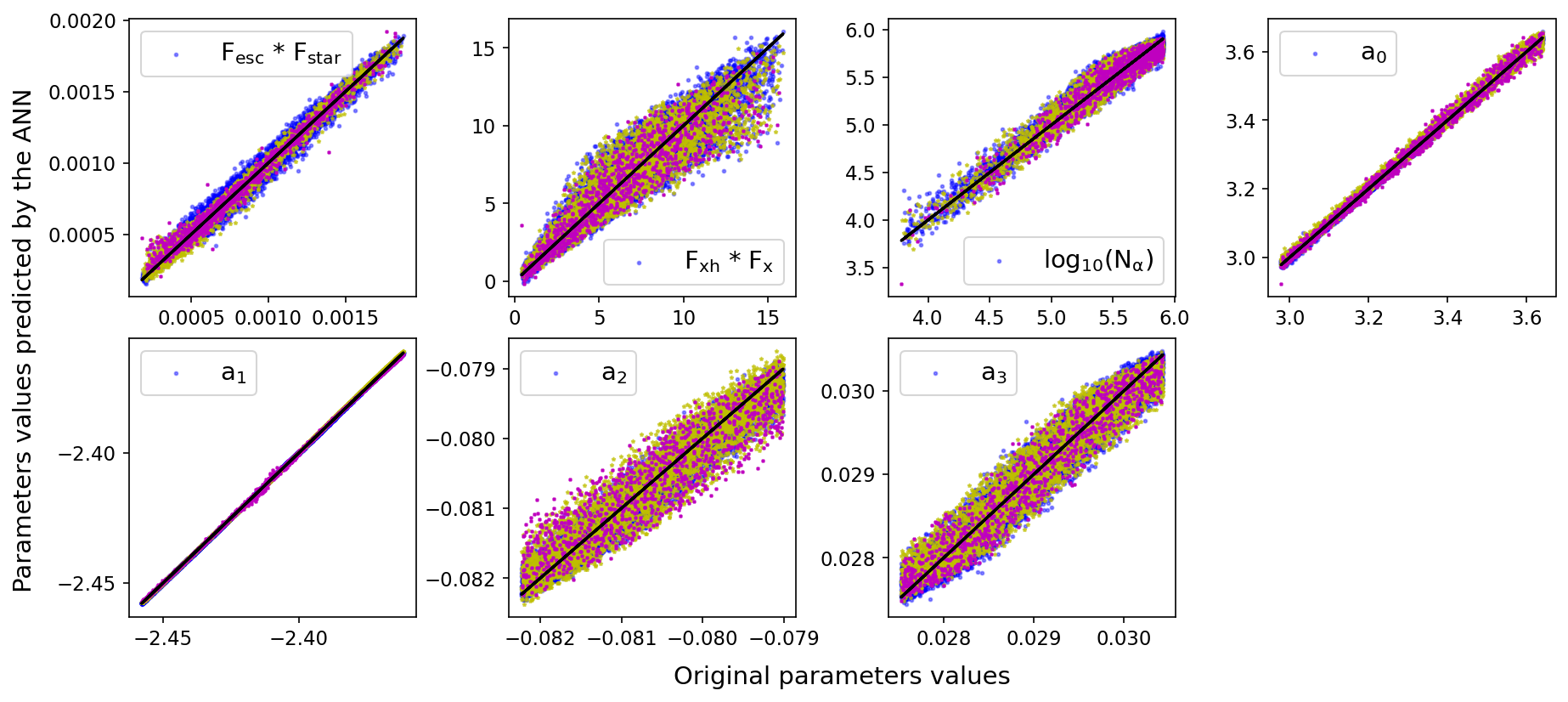}}
    
    \caption{The scatter plots above show predicted signal and foreground parameter values obtained through an ANN model trained on a physical global 21cm signal. The signal and foreground data sets are generated by sampling the parameter space using Hammersley sequence sampling in three sizes: 10,000, 50,000, and 100,000 samples. Magenta scatter points in each scatter plot denote predictions made by the ANN trained with 10,000 samples, while yellow and blue scatter points indicate predictions from ANN models trained with 50,000 and 100,000 samples, respectively. The actual values of the parameters are plotted in a solid black line in each plot.}
   
    \label{Fig12}
\end{figure*}

\begin{figure*}
    \centering
    {\includegraphics[width=1.0\textwidth]{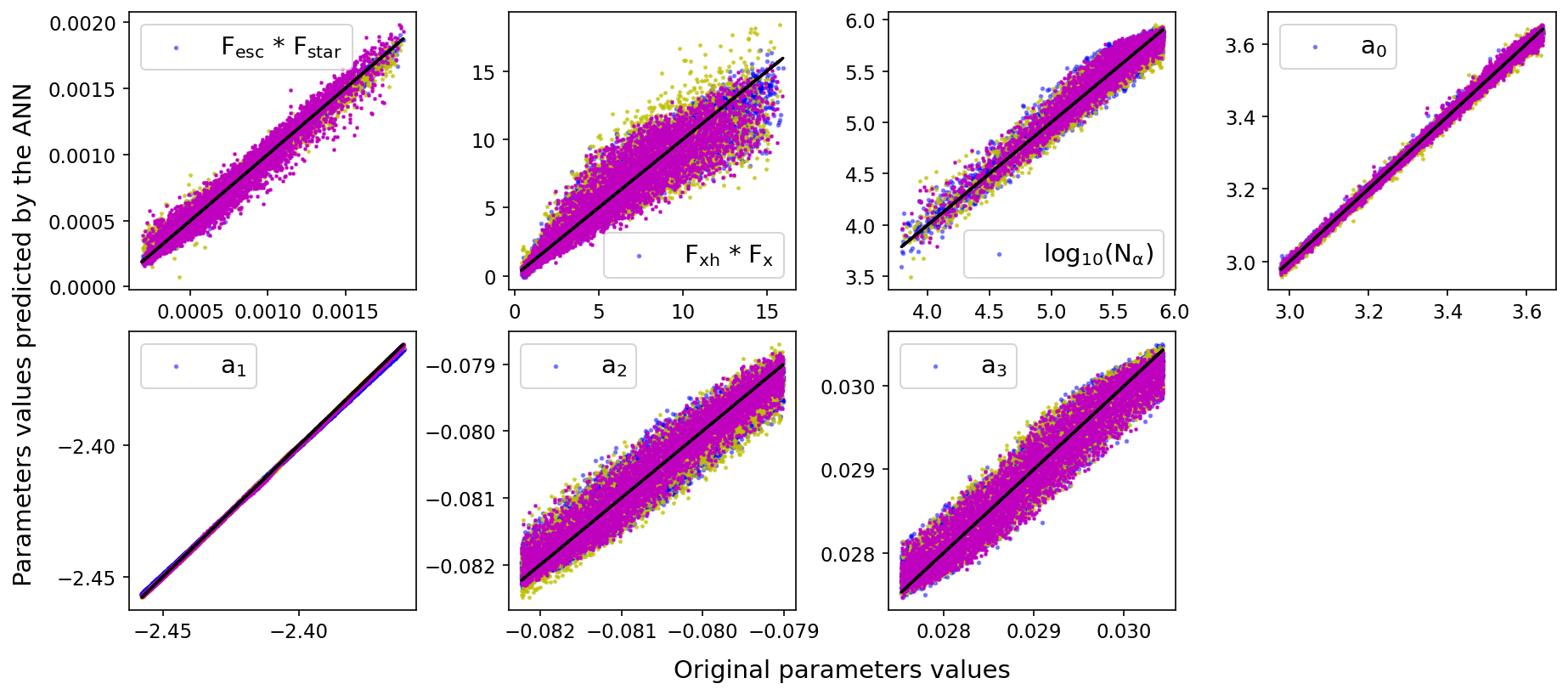}}
    
    \caption{The scatter plots above show predicted signal and foreground parameter values obtained through an ANN model trained on a physical global 21cm signal. The signal and foreground data sets are generated by sampling the parameter space using Latin hypercube sampling in three sizes: 10,000, 50,000, and 200,000 samples. Magenta scatter points in each scatter plot denote predictions made by the ANN trained with 10,000 samples, while yellow and blue scatter points indicate predictions from ANN models trained with 50,000 and 100,000 samples, respectively. The actual values of the parameters are plotted in a solid black line in each plot.}
   
    \label{Fig13}
\end{figure*}

\begin{figure*}
    \centering
    {\includegraphics[width=0.99\textwidth]{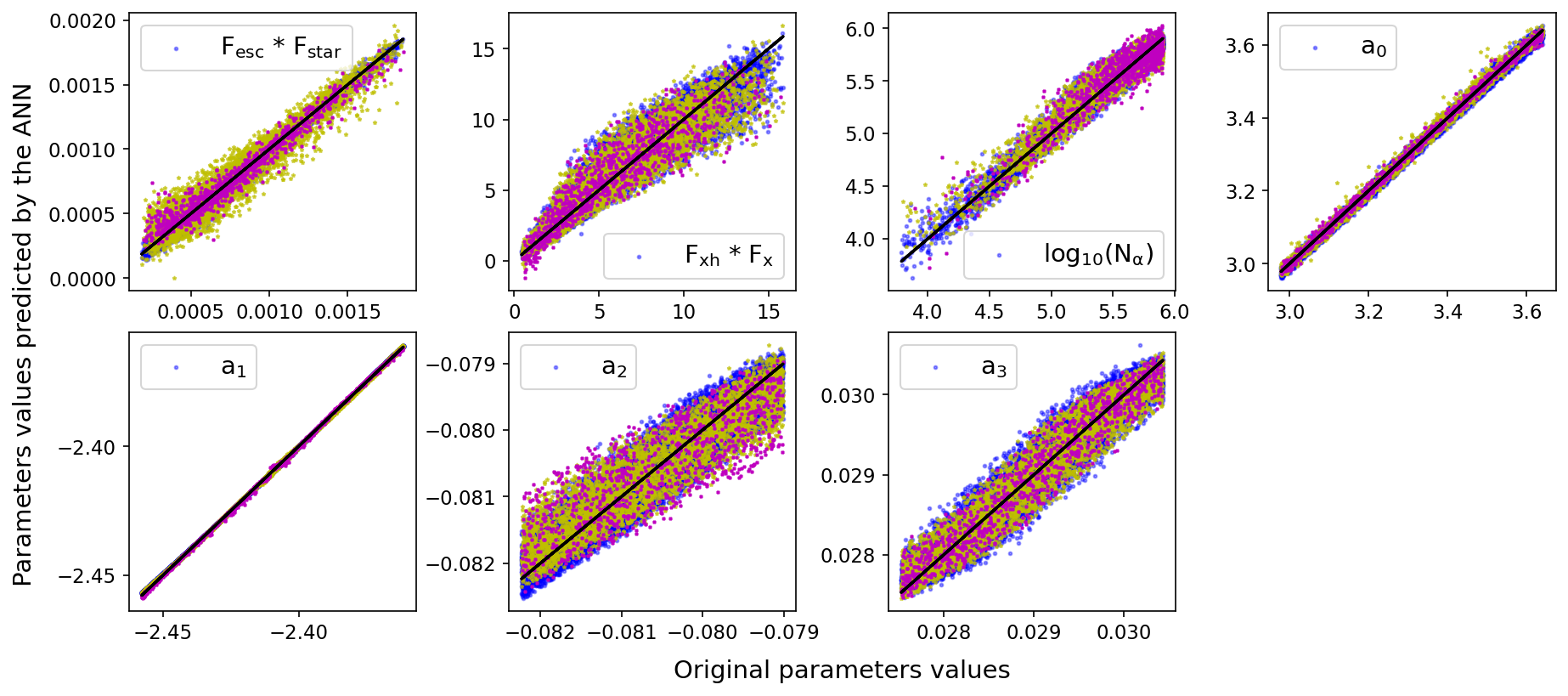}}
    
    \caption{The scatter plots above show predicted signal and foreground parameter values obtained through an ANN model trained on a physical global 21cm signal. The signal and foreground data sets are generated by sampling the parameter space using Random sampling in three sizes: 10,000, 50,000, and 100,000 samples. Magenta scatter points in each scatter plot denote predictions made by the ANN trained with 10,000 samples, while yellow and blue scatter points indicate predictions from ANN models trained with 50,000 and 100,000 samples, respectively. The actual values of the parameters are plotted in a solid black line in each plot.}
   
    \label{Fig14}
\end{figure*}

Similar to the parametrized  case, we introduced the effects of foreground and thermal noise in the non-parametric signal. Given fewer free parameters than the parametric case, we generated datasets in three sizes: 10,000, 50,000, and 100,000 samples. The architecture of the ANN model closely resembles that of the parametrized  case, with the primary difference being the output layer of the second ANN model, which comprises 5 neurons. Each neuron represents 3 signal parameters and 2 foreground parameters ($a_{0}$, $a_{3}$). Here, we observe a consistent trend similar to parametric scenarios: the ANN model trained with 10000 datasets achieves an overall $\rm R^{2}$ score of around 0.92, which increases to approximately 0.94 with 50000 datasets, and around 0.96 with 10,0000 datasets. The detailed results for each sampling method with the various dataset sizes, the $ \rm R^2$ and RMSE score for individual parameters, are listed in Tab. \ref{tab11} and Tab. \ref{tab12}. We also individually visualized the predicted parameter values against the original values for each sampling method across different dataset sizes. These visualizations are presented in Fig. \ref{Fig12} for HSS, Fig. \ref{Fig13} for LHS, and Fig. \ref{Fig14} for Random sampling. 

In our study, we observed that for optimal performance of the ANN model with any sampling method, training with a sufficient number of datasets is essential. For example, training the ANN model with 100,000 datasets resulted in precise prediction of signal parameters, with $\rm R^{2}$ scores ranging from 0.92 to 0.98 and root mean square error (RMSE) values between 0.021 and 0.066. Additionally, the ANN effectively predicted foreground parameters, yielding $\rm R^{2}$ scores ranging from 0.95 to 0.99 and RMSE values between 0.008 and 0.064, showcasing significantly improved accuracy compared to the model trained with 10,000 datasets. Detailed results are provided in Tab.\ref{tab11} and Tab.\ref{tab12}.

\begin{table*}
\centering
\begin{tabular}{|l|l|l|l|l|l|l|l|l|l|}
\hline
Size        &        & 10000     &         &         &50000 &        &     &100000 &             \\ \hline
                    & HSS      & LHS     & Rand   & HSS     & LHS   & Rand      & HSS       & LHS  & Rand   \\  \hline
Avg.   & 0.9277  &0.9187   & 0.9018   & 0.9505 & 0.9448     & 0.9429 & 0.9670 & 0.9647 & 0.9608 \\ \hline
$f_{\ast}.f_{esc}$  & 0.9813  & 0.9532  & 0.9776   & 0.9890      & 0.9510     & 0.9634       & 0.9843   & 0.9903  & 0.9860 \\
$f_{X,h}.f_{X}$     & 0.8574  & 0.8727  & 0.8272   & 0.8807      & 0.8502     & 0.8570       & 0.9139   & 0.9168  & 0.9142  \\
$N_{\alpha}$        & 0.9395  & 0.9398  & 0.8502   & 0.9512      & 0.9454     & 0.9405       & 0.9620   & 0.9563  & 0.9629 \\  
$\rm a_{0}$         & 0.9925  & 0.9978  & 0.9960   & 0.9968      & 0.9959     & 0.9969       & 0.9988   & 0.9986  & 0.9973  \\
$\rm a_{1}$         & 0.9996  & 0.9969  & 0.9974   & 0.9993      & 0.9982     & 0.9998       & 0.9999   & 0.9991  & 0.9998  \\
$\rm a_{2}$         & 0.7921  & 0.7281  & 0.7291   & 0.9080      & 0.9386     & 0.9084       & 0.9590   & 0.9415  & 0.9242   \\
$\rm a_{3}$         & 0.9317  & 0.9427  & 0.9353   & 0.9340      & 0.9327     & 0.9337       & 0.9505   & 0.9504  & 0.9400    \\ \hline

\end{tabular}
\caption{The computed $\rm R^2$-scores for all signal and foreground parameters for each case studied are listed here. We used the physical model to construct the global 21cm signal and the log-log polynomial to construct the foreground.}
\label{tab11}
\end{table*}

\begin{table*}
\centering
\begin{tabular}{|l|l|l|l|l|l|l|l|l|l|}
\hline
Size        &        & 10000     &         &         &50000   &        &     &100000 &  \\ \hline
                     & HSS     & LHS     & Rand     & HSS     & LHS    & Rand   & HSS     & LHS     & Rand   \\  \hline
$f_{\ast}.f_{esc}$  & 0.0303  & 0.0457   & 0.0319   & 0.0219  & 0.0460 & 0.0403  & 0.0265  & 0.0205  & 0.0246 \\
$f_{X,h}.f_{X}$     & 0.0886  & 0.0819   & 0.0945   & 0.0770  & 0.0885 & 0.0853  & 0.0670  & 0.0661  & 0.0675  \\
$N_{\alpha}$        & 0.0457  & 0.0433   & 0.0715   & 0.0424  & 0.0432 & 0.0456  & 0.0363  & 0.0386  & 0.0361 \\   
$\rm a_{0}$         & 0.0248  & 0.0137   & 0.0179   & 0.0113  & 0.0184 & 0.0151  & 0.0098  & 0.0107  & 0.0149\\
$\rm a_{1}$         & 0.0054  & 0.0159   & 0.0145   & 0.0075  & 0.0037 & 0.0036  & 0.0023  & 0.0085  & 0.0035\\
$\rm a_{2}$         & 0.1279  & 0.1497   & 0.1465   & 0.0880  & 0.0714 & 0.0878  & 0.0581  & 0.0695  & 0.0763\\
$\rm a_{3}$         & 0.0757  & 0.0708   & 0.0739   & 0.0741  & 0.0749 & 0.0740  & 0.0644  & 0.0643  & 0.0711\\ \hline
\end{tabular}
\caption{The computed $\rm RMSE $-scores for all signal and foreground parameters for each case studied are listed here. We used the physical model to construct the global 21cm signal and the log-log polynomial to construct the foreground.}
\label{tab12}
\end{table*}

\subsection{Generalizability test} \label{appendix_generalizability}
\begin{figure*}
    \centering
    {\includegraphics[width=0.9\textwidth]{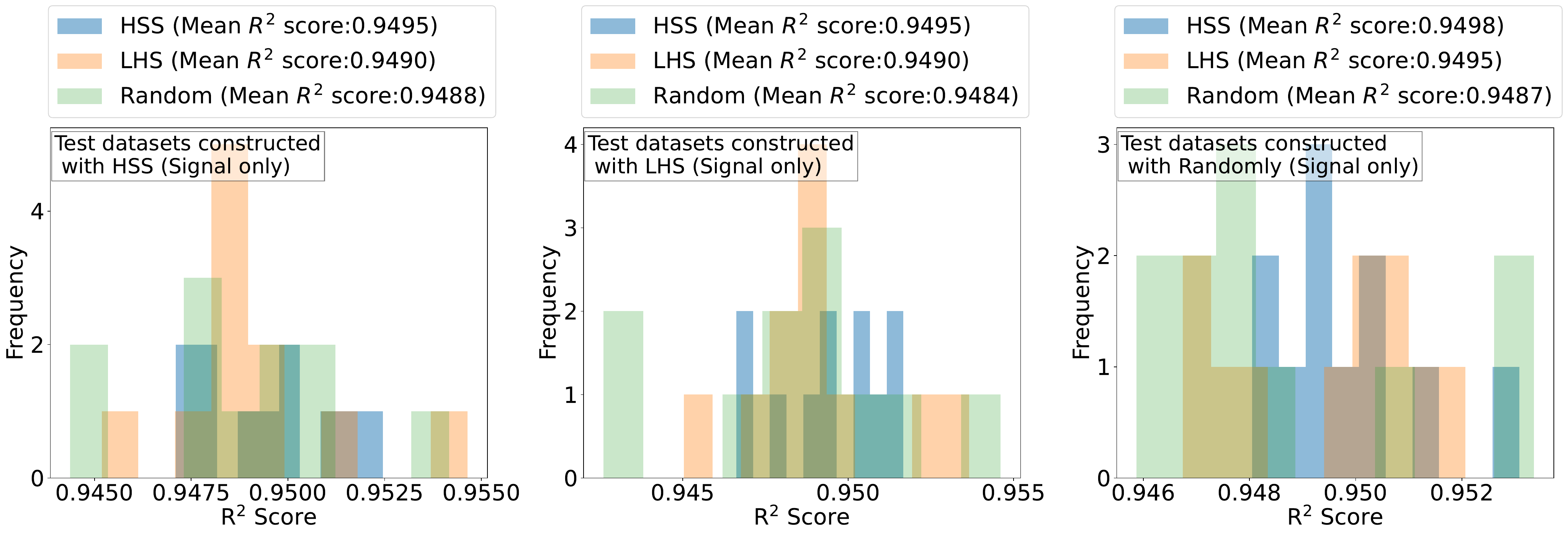}}
    {\includegraphics[width=0.9\textwidth]{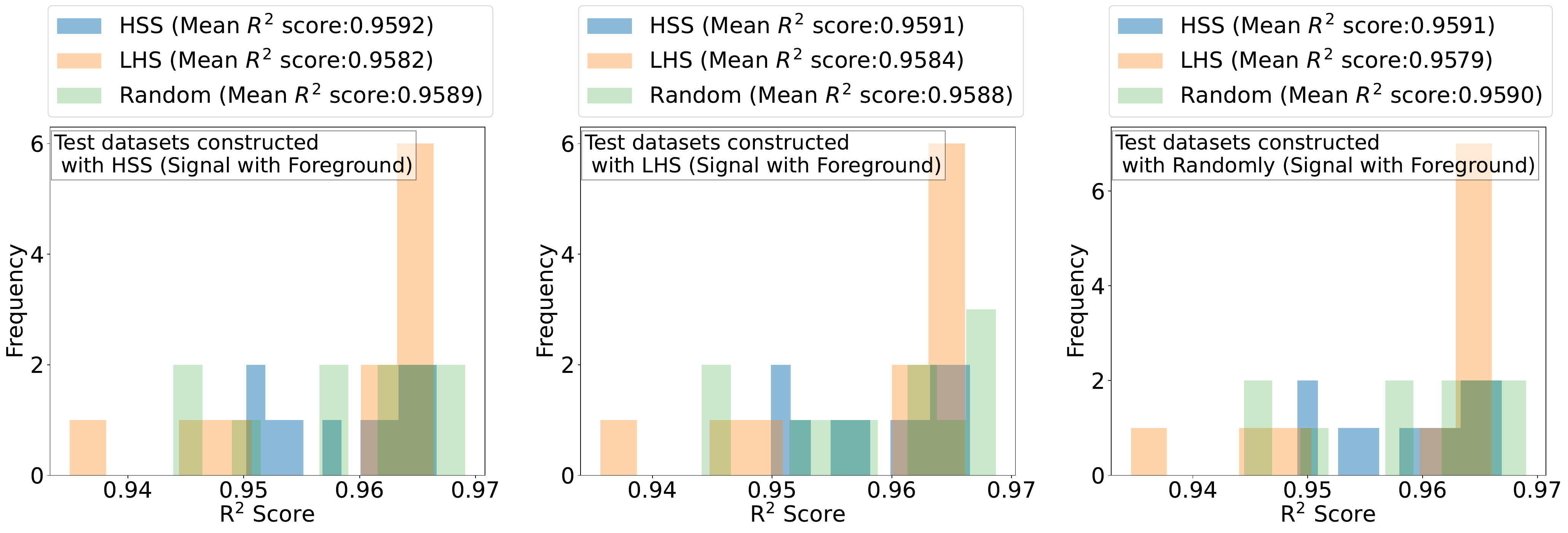}}
    \caption{The figure illustrates ANN model predictions for various trials trained on datasets sampled using HSS, LHS, and Random methods with optimal sample sizes. Histograms depict ANN prediction accuracy measured by $\rm R^{2}$ scores. Panels in the figure's rows correspond to different scenarios: the top three panels in the first row show predictions for the Non-parametrized  signal, followed by predictions for the Non-parametrized  signal with foreground and thermal noise in the second row. Histograms are colour-coded: blue represents HSS-trained ANN models, orange represents LHS-trained ANN models, and green represents the Random-trained ANN model's prediction accuracy. }
   
    \label{Fig_robustness_Non_para}
\end{figure*}

We have conducted a generalizability test to demonstrate the robustness of the ANN model for non-parametrized  scenarios. The detailed results are plotted below in the histogram, see Fig. \ref{Fig_robustness_Non_para}. 

\section{Training loss and validation loss of the ANN models}
\begin{figure*}
    \centering
    {\includegraphics[width=0.45\textwidth]{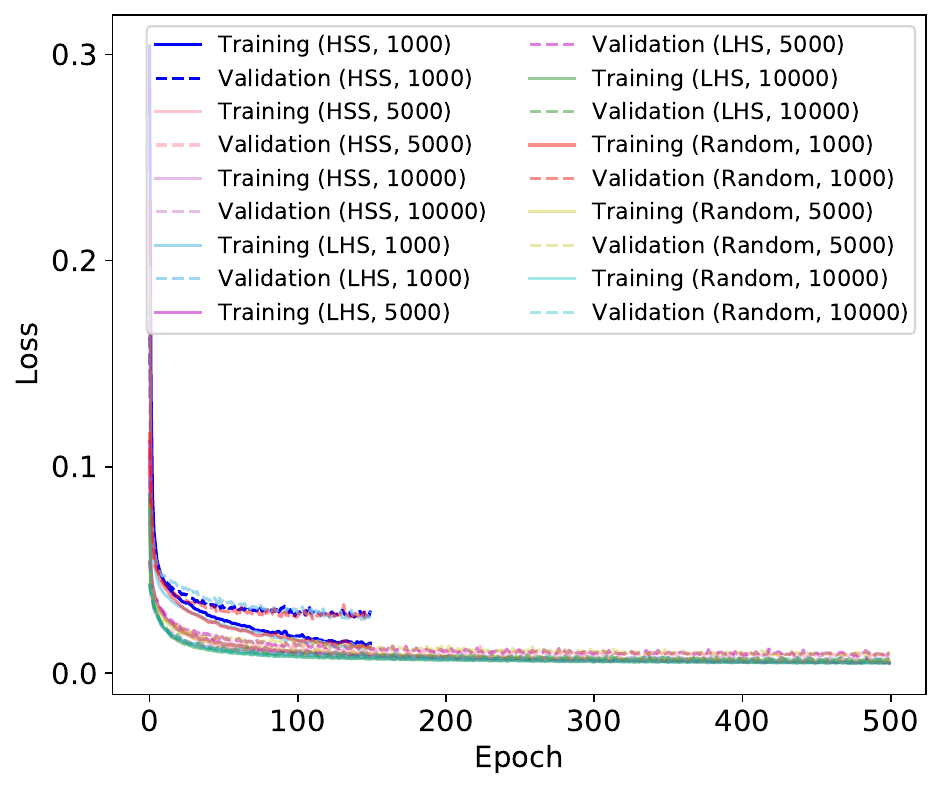}}
    {\includegraphics[width=0.45\textwidth]{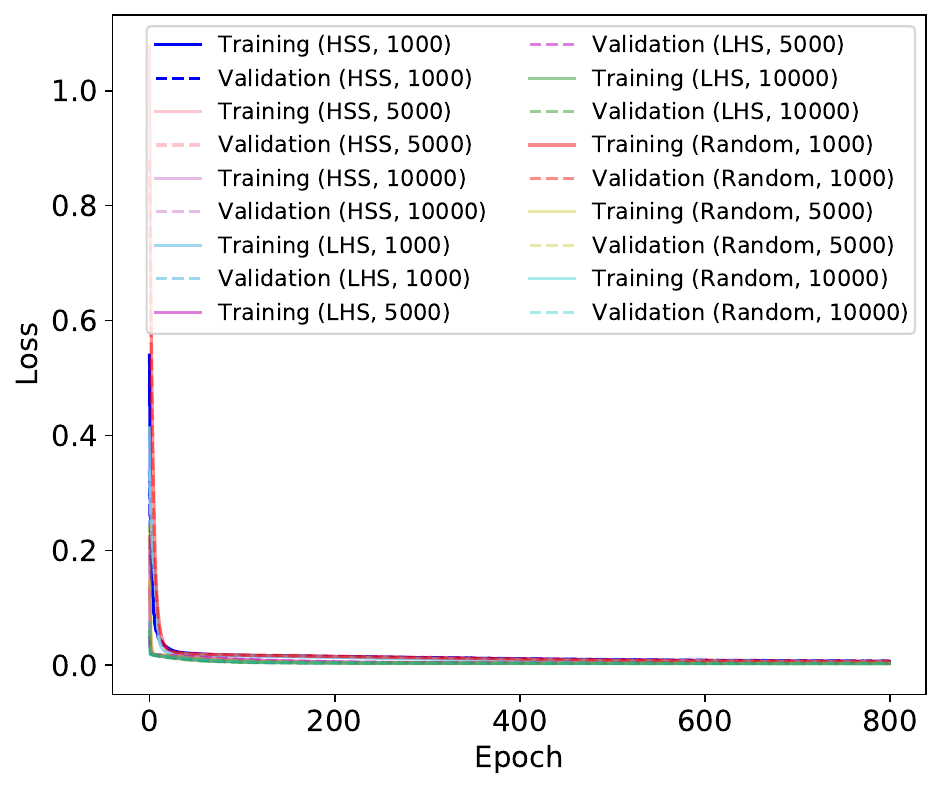}}
    {\includegraphics[width=0.45\textwidth]{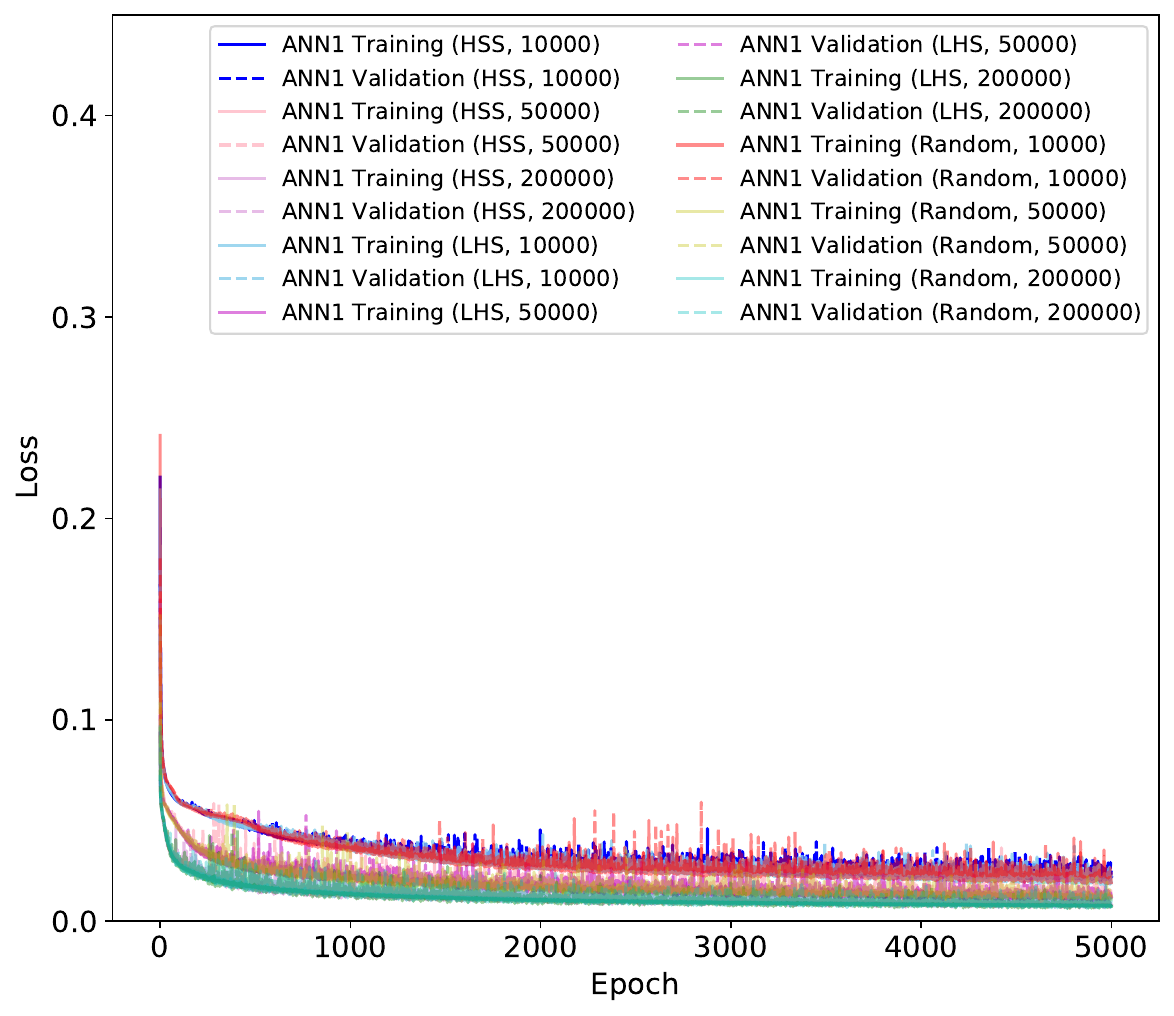}}
    {\includegraphics[width=0.45\textwidth]{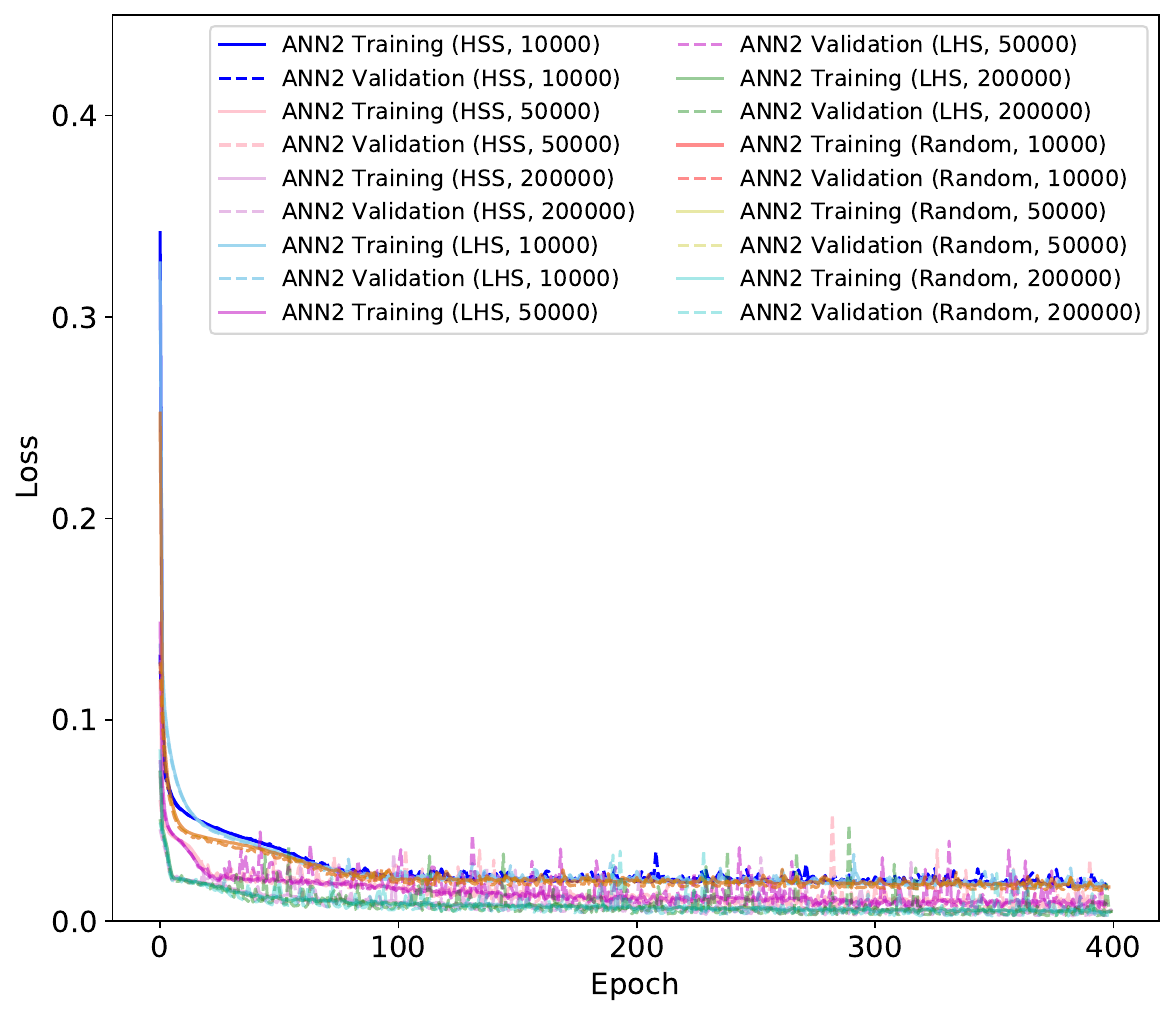}}
    {\includegraphics[width=0.45\textwidth]{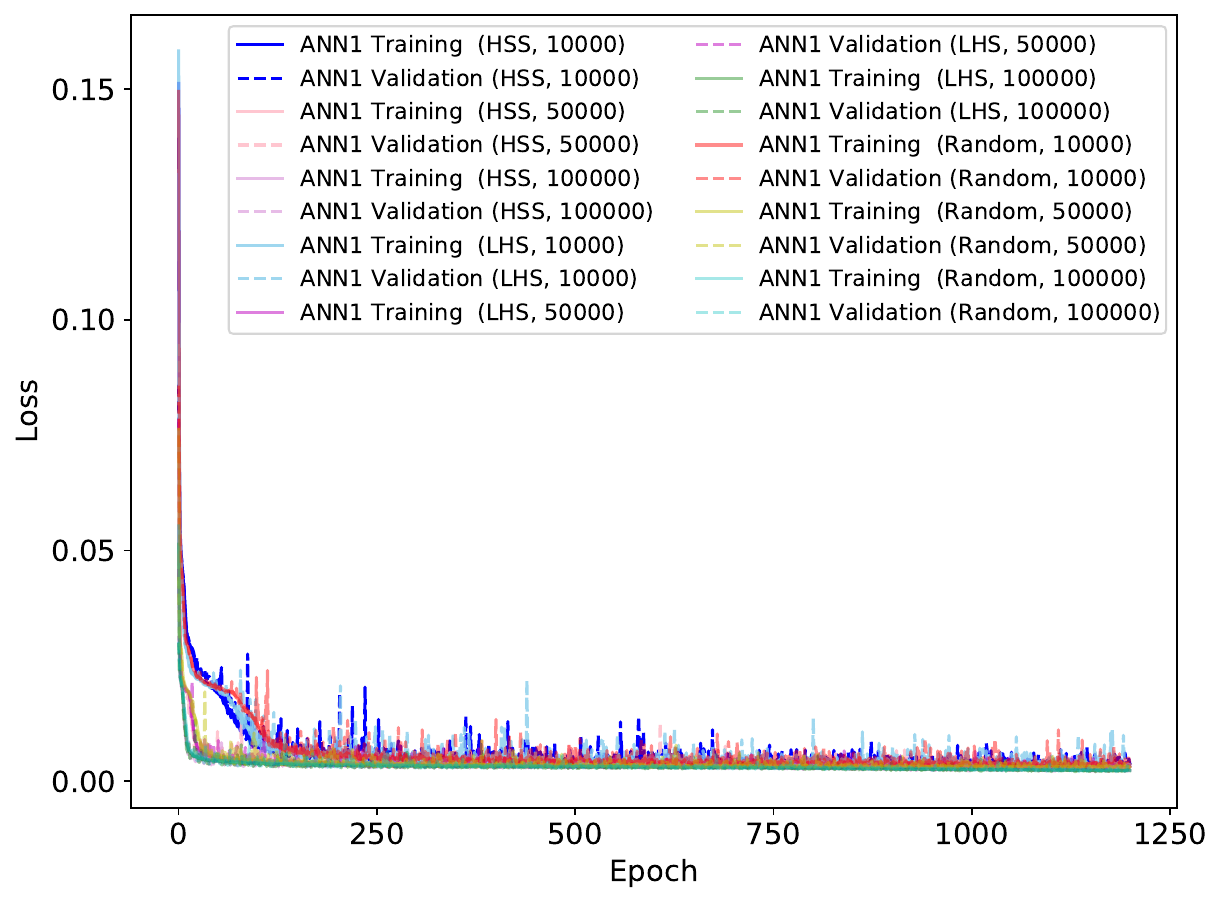}}
    {\includegraphics[width=0.45\textwidth]{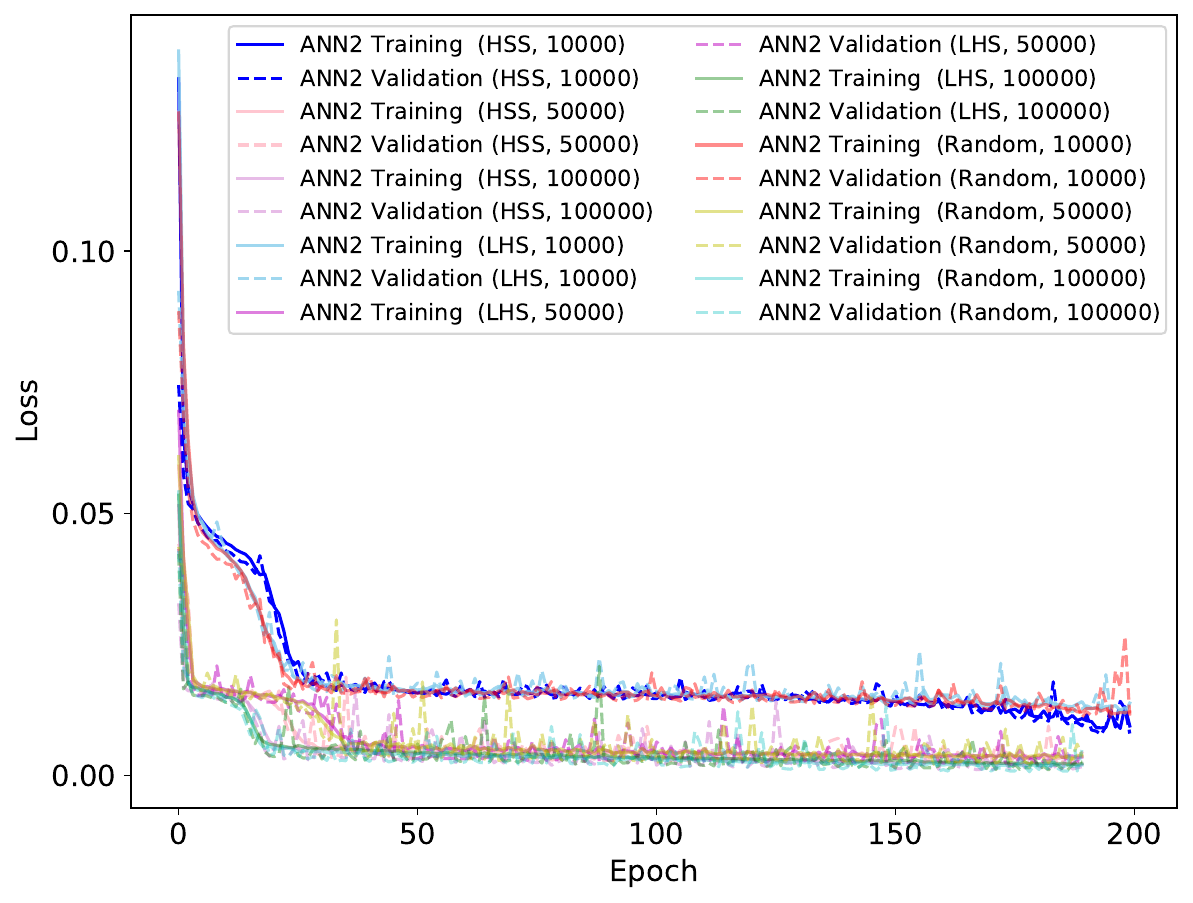}}
    \caption{The figure illustrates the evolution of the network's loss function across various scenarios. The training loss is denoted by a solid line, and the validation loss is indicated by a dashed line over epochs. Notably, the test loss closely follows the training loss in this visualization. \textbf{Top row:} training loss for parametrized  signal (\textbf{left}) and non-parametrized  signal (\textbf{right}). \textbf{Second row}: training and validation loss for parametrized  signals with foreground. \textbf{Bottom row}: training and validation loss for foreground-added non-parametrized  signals.}
   
    \label{Fig_loss}
\end{figure*}

We have plotted the training and validation loss of the ANN models for various scenarios; see Fig.\ref{Fig_loss}. 

\bibliographystyle{JHEP}
\bibliography{biblio}
\end{document}